\pdfoutput=1
\documentclass[journal]{IEEEtran}

\usepackage[T1]{fontenc}
\usepackage{amsmath,amssymb,amsfonts}
\usepackage{graphicx}
\usepackage{booktabs}
\usepackage{multirow}
\usepackage{array}
\usepackage{url}
\usepackage{flafter} 

\DeclareMathOperator{\Mean}{Mean}

\DeclareMathOperator{\MHA}{MHA}
\DeclareMathOperator{\LN}{LN}
\DeclareMathOperator{\MLPop}{MLP}
\newcommand{\CN}{\mathcal{CN}}

\newtheorem{remark}{Remark}

\newcommand{\noheadpunct}[1]{\ \,}
\begin{document}

\title{LWM-TF: A Wireless Foundation Model on\\ 5G NR Time-Frequency Resource Grids}

\author{Namhyun~Kim, Sadjad~Alikhani, and Ahmed~Alkhateeb%
\thanks{The authors are with the School of Electrical, Computer and Energy
Engineering, Arizona State University, Tempe, AZ 85287 USA (e-mail:
\{namhyunk, alikhani, alkhateeb\}@asu.edu).}}

\maketitle

\begin{abstract}
Received OFDM grids preserve both pilot and payload observations, yet wireless
foundation models are rarely pre-trained on them. We introduce LWM-TF, a
9.58\,M-parameter factorized masked autoencoder pre-trained on complex 5G NR
uplink grids taken after the FFT and before channel estimation. Each block
attends first across OFDM symbols and then across resource blocks and receive
antennas, and the mask follows the same two axes. Because the payload phase is
unpredictable, the model reconstructs the normalized log-power of each masked
token and predicts its scale separately; ablations show that the two terms are
complementary and that this target outperforms complex-signal reconstruction.
The encoder is pre-trained on $850{,}496$ physical uplink shared channel
(PUSCH) slots from a $10^6$-slot dataset simulated with the NR transmit chain
under various channel conditions. With the encoder
frozen, linear probes classify modulation, predict CRC success, and detect UE
mobility. On a separately generated test set, $16$ labeled slots per class
give $0.85$ modulation accuracy, $0.82$ CRC area under the ROC curve (AUROC),
and $0.73$ mobility AUROC, ahead of a supervised CNN on every paired draw.
With $10{,}000$ labels the probes reach $0.94$, $0.96$, and $0.92$, well
above engineered receiver statistics, and detect mobility where the CNN stays
at chance. Without any impairment labels, distances in the frozen feature
space flag unseen waveform impairments with $0.78$ macro AUROC, against
$0.68$ for receiver statistics. A single pre-trained encoder thus provides
receiver features that transfer across tasks, most strongly when labels are
scarce.
\end{abstract}

\begin{IEEEkeywords}
5G NR PUSCH, masked autoencoder, self-supervised learning, time-frequency
resource grid, wireless foundation model, few-shot learning, CRC decodability,
modulation classification, mobility detection.
\end{IEEEkeywords}

\IEEEpeerreviewmaketitle

\section{Introduction}
\label{sec:intro}
The sixth generation (6G) of wireless networks is widely expected to embed
machine learning across all layers of the radio stack rather than treat it as
an auxiliary tool \cite{chen2024bigai}. At the
physical layer (PHY), however, most learning-based solutions are still
developed for individual tasks, with separate models, supervision, and
optimization pipelines \cite{oshea2017intro}--\cite{aitaoudia2022e2e}. This
task-specific paradigm limits the reuse of learned knowledge and makes each
new receiver function dependent on another labeled dataset and training cycle.
Foundation models offer a different path: a high-capacity encoder can first
learn broadly useful representations from large amounts of unlabeled data and
then be adapted to multiple downstream tasks through lightweight task-specific
heads, as demonstrated in language, vision, and speech
\cite{devlin2019bert,brown2020gpt3,he2022mae,baevski2020wav2vec2}. Extending
this paradigm to the wireless PHY could therefore turn costly per-task model
development into the reuse of a shared representation.

Wireless foundation models (WFMs) pre-train on channel coefficients, impulse
responses, or pilot observations~\cite{alikhani2025lwm,liu2025wifo,guler2026contrawimae,guler2026pilotwimae,ott2024radiofm},
magnitude spectrograms~\cite{aboulfotouh2024radio,kim2026lwmspectro},
or raw in-phase and quadrature (I/Q) streams~\cite{mashaal2026iqfm};
multi-modal models combine these views~\cite{aboulfotouh2025wavesfm,mohamed2026latentwave}.
For post-FFT receiver learning, each input has a limitation: channel and pilot
data omit payload-bearing observations, magnitude spectrograms
discard phase (including across antennas), and raw I/Q leaves the
resource-element layout implicit. The received time-frequency grid preserves
complex pilot and payload observations with channel and noise effects, indexed
by symbol, subcarrier, and antenna. It thus combines the signal content
and structure suited for a shared receiver representation.

\begin{figure}[!t]
\centering
\includegraphics[width=\linewidth]{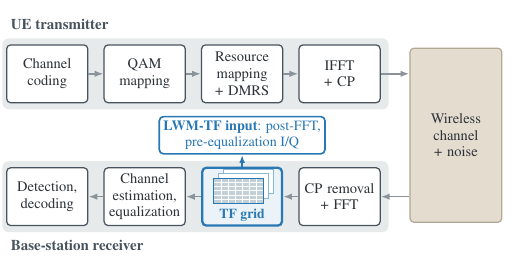}
\caption{Uplink OFDM link. The LWM-TF input is the received time-frequency (TF)
grid after cyclic prefix (CP) removal and the FFT, before channel estimation.}
\label{fig:rxchain}
\end{figure}

Accordingly, we pre-train directly on the received orthogonal frequency-division
multiplexing (OFDM) grid, taken after the fast Fourier transform (FFT) and
before channel estimation (Fig.~\ref{fig:rxchain}), so that the
representation does not inherit a channel estimator's assumptions or errors
(Section~\ref{sec:crclabel}). To our knowledge, no WFM has pre-trained on this
grid, whose payload phase is unpredictable and whose receiver labels require a
full transmit and receive chain;
WavesFM~\cite{aboulfotouh2025wavesfm} uses received OFDM pilots for
channel-estimation fine-tuning. {We run this chain in a
3GPP-compliant physical uplink shared channel (PUSCH) simulator: the transmit chain follows
TS~38.211/38.212~\cite{3gpp38211,3gpp38212} with the modulation and coding scheme (MCS) tables of
TS~38.214~\cite{3gpp38214}, propagation follows the TR~38.901~\cite{3gpp38901}
urban macro (UMa), urban micro (UMi), and rural macro (RMa) models, and the
cyclic redundancy check (CRC) labels come from a standard receiver
(Section~\ref{sec:crclabel}). Every input, and every label except the user equipment (UE) speed behind the
mobility label, is therefore a quantity that a 5G New Radio (NR) base station
handles.}

On this grid, LWM-TF adapts masked reconstruction~\cite{he2022mae} to the
signal's phase structure: because the per-element data phase carries the
scrambled payload, the model reconstructs masked, patch-normalized log-power and
predicts each patch's scale. {In downstream evaluations, probes
using the frozen encoder outperform engineered-statistic and
random-initialization baselines on modulation recognition, CRC decodability, and
channel time-variation (mobility) detection. Against the evaluated
convolutional neural network (CNN),
they improve mobility with all labels and all three tasks with a few labels
per class.}

\subsection{Related Work}

\textbf{Learned receivers and task-specific PHY models.}
Deep learning has been applied to virtually every block of the wireless
receive chain~\cite{oshea2017intro}. Several of these models already consume
the received OFDM grid: DeepRx~\cite{honkala2021deeprx} maps the post-FFT grid,
together with pilot information, to bit log-likelihood ratios, and deep
networks estimate channels and detect symbols from received OFDM
symbols~\cite{ye2018power}. Others classify modulation from raw
I/Q~\cite{oshea2018ota}, compress channel state information (CSI) feedback~\cite{wen2018csinet}, manage
beams in high-mobility millimeter-wave (mmWave) systems~\cite{alkhateeb2018coordinated}, or learn
the air interface end to end~\cite{aitaoudia2022e2e}. Each is trained with its
own labels for one task: the received grid is already used as an input, but no
representation of it is shared across tasks.

\textbf{Wireless foundation models.}
{
Wireless foundation models can be grouped into channel-based, signal-based,
and multi-input approaches (Table~\ref{tab:wfm}); the families overlap, and the
learning methods listed for each are alternatives, not requirements.

\emph{Channel-based approaches} learn propagation structure from channel
coefficients or pilot observations. LWM~\cite{alikhani2025lwm} reconstructs randomly masked patches
of DeepMIMO channels and evaluates beam prediction and line-of-sight (LoS)/non-LoS (NLoS) classification.
WiFo~\cite{liu2025wifo} uses simulated space--time--frequency channels and
random, temporal, and frequency masks for channel prediction.
ContraWiMAE~\cite{guler2026contrawimae} combines masked reconstruction with
contrastive learning for channel estimation, beam selection, and LoS detection.
PilotWiMAE~\cite{guler2026pilotwimae} encodes noisy pilot observations with
attention factorized into time and space--frequency and pairs patch-normalized
reconstruction with a scale loss.
Ott et al.~\cite{ott2024radiofm} reconstruct masked 5G channel impulse responses
for indoor localization, WiFo-CF~\cite{liu2026wifocf} targets CSI feedback, and
LLM4CP~\cite{liu2024llm4cp} adapts a pre-trained language model to channel
prediction.

\emph{Signal-based approaches} learn from received observations rather than
channel coefficients. Spectrogram models~\cite{aboulfotouh2024radio,kim2026lwmspectro}
reconstruct masked time--frequency patches for signal classification and
sensing. LWM-Spectro uses simulated WiFi, LTE, and 5G signals with
DeepMIMO-based channels, adds supervised contrastive learning during
adaptation, and evaluates modulation and SNR/mobility classification.
IQFM~\cite{mashaal2026iqfm} instead applies contrastive learning to measured
I/Q streams, with evaluations including modulation, arrival angle, beam
prediction, and radio-frequency (RF) fingerprinting.

\emph{Multi-input approaches} extend a shared model to different wireless
representations. WavesFM~\cite{aboulfotouh2025wavesfm} pre-trains through
uniformly masked reconstruction of measured spectrograms and CSI, supports
pilot I/Q on the OFDM grid in fine-tuning, and adapts to sensing, positioning, RF classification, and
channel estimation. LatentWave~\cite{mohamed2026latentwave} learns from
spectrograms and CSI by predicting masked features in latent space.

LWM-TF belongs to the signal-based family. Unlike the pilot-only PilotWiMAE,
it uses all complex PUSCH samples, pilot and payload, arranged by OFDM symbol,
subcarrier, and receive antenna, before channel estimation and equalization. Pre-training combines log-power reconstruction
with patch-scale supervision; the main model uses uniform masking, with
masking stratified by demodulation reference signal (DMRS) position evaluated
separately. A frozen encoder is evaluated
on modulation classification, CRC decodability, channel time-variation
detection, SNR estimation (as a control), and waveform anomaly detection.
LWM-TF uses a separately generated Sionna PUSCH dataset and does not reuse
the datasets, signal-preprocessing pipelines, or pre-trained weights of
LWM, LWM-Spectro, or WavesFM, which pre-train on channel matrices,
spectrograms, and CSI rather than received PUSCH grids, so their weights
expect other inputs and patch shapes. The comparison describes the published
approaches, not their relative performance on a common benchmark.
}

\textbf{Self-supervised objectives.}
Foundation models in language, speech, and vision are pre-trained with three
families of objectives: masked prediction of the
input~\cite{devlin2019bert,he2022mae}, contrastive
objectives~\cite{chen2020simclr,baevski2020wav2vec2}, and prediction of masked
content in a learned latent space~\cite{assran2023ijepa}. Wireless foundation
models draw on all three: masked reconstruction of channels or spectrograms
(LWM, WiFo, LWM-Spectro), contrastive learning (ContraWiMAE, IQFM), and latent
prediction (LatentWave). LWM-TF follows the masked
autoencoder~\cite{he2022mae}, whose asymmetric design encodes only the visible
tokens and reconstructs the rest with a light decoder, keeping a $75\%$ masking
ratio affordable; it departs by choosing the reconstruction target
from the structure of the resource grid and by factorizing the mask along the
symbol and antenna--physical resource block (PRB) axes.

\subsection{Contributions}
\label{sec:contrib}
The main contributions are:
\begin{enumerate}
\item {\textbf{A factorized masked autoencoder for the 5G NR receiver grid.}
LWM-TF encodes the complex received grid, with one token per PRB, OFDM
symbol, and receive antenna; its attention and mask are factorized along the
symbol and antenna--PRB axes of the grid. It reconstructs
patch-normalized log-power and predicts per-patch scale. Since fully masked
payload symbols are unpredictable, complex-value reconstruction provides no
useful average target for them. Controlled ablations show that log-power
reconstruction with the scale target outperforms complex reconstruction.}

\item {\textbf{A 3GPP-compliant dataset and an independent test set.}
We generate a $10^6$-slot single-input multiple-output (SIMO) PUSCH dataset whose transmit chain follows
TS~38.211/38.212 and whose channels follow the TR~38.901 UMa, UMi, and RMa
models and DeepMIMO channels, using target signal-to-noise ratio (SNR) calibration and modulation-balanced sampling. Grouped data splits
and three training seeds support controlled comparisons. A second test
set of $24{,}576$ slots was generated after the original 64 evaluation
candidates were frozen. The extended main model was chosen using development
validation only; no model choice depends on the second test set.}

\item {\textbf{Label-efficient payload, link, and channel tasks from one frozen encoder.}
On the second test set, frozen linear probes reach $0.939$ modulation accuracy
and $0.964$ CRC area under the receiver operating characteristic curve
(AUROC) with $10{,}000$ labels. With $100$ labels they reach
$0.87$ accuracy and $0.89$ AUROC; on channel time-variation (mobility)
detection they reach $0.89$ AUROC with $1{,}000$. The comparisons use
three training seeds and two separately generated test sets.}
\end{enumerate}

\textbf{Notation.} Bold lowercase letters denote vectors and bold uppercase
letters denote matrices and tensors; overbars denote pooled (typically
mean-pooled) representations. Complex conjugation is written
$(\cdot)^*$, the expectation operator $\mathbb{E}[\cdot]$, the
absolute value $|\cdot|$, and $\CN(0,\sigma^2\mathbf{I})$ a circularly
symmetric complex Gaussian. Classification uses accuracy; regression uses
$R^2=1-\sum_i(y_i-\hat y_i)^2/\sum_i(y_i-\bar y)^2$ for test targets $y_i$ with
mean $\bar y$ and predictions $\hat y_i$ ($1$ is perfect, $0$ matches predicting
$\bar y$, and negative values are worse). Binary AUROC is
\begin{equation*}
\mathrm{AUROC}=\frac{1}{N_+N_-}\sum_{i,j}
\left(\mathbb{I}[s_i^+>s_j^-]+\tfrac12\mathbb{I}[s_i^+=s_j^-]\right),
\end{equation*}
where $s_i^+,s_j^-$ are the scores of the $N_+$ positive and $N_-$ negative
test samples and $\mathbb{I}$ is the indicator; ties count one half, and $0.5$
is chance.



\begin{table*}[!t]
\caption{Families of wireless foundation models.}
\label{tab:wfm}
\centering
\footnotesize
\begin{tabular}{@{}>{\raggedright\arraybackslash}p{0.13\linewidth}>{\raggedright\arraybackslash}p{0.22\linewidth}>{\raggedright\arraybackslash}p{0.28\linewidth}>{\raggedright\arraybackslash}p{0.27\linewidth}@{}}
\hline
\rule{0pt}{8.2pt}\textbf{Approach} & \textbf{Input} & \textbf{Learning} & \textbf{Studies}\\
\hline
Channel & Channel coefficients, impulse responses, or pilot observations over antennas, frequency, and time & Masked channel reconstruction; contrastive learning; language-model adaptation & LWM~\cite{alikhani2025lwm}, WiFo~\cite{liu2025wifo}, ContraWiMAE~\cite{guler2026contrawimae}, PilotWiMAE~\cite{guler2026pilotwimae}, Ott et al.~\cite{ott2024radiofm}, WiFo-CF~\cite{liu2026wifocf}, LLM4CP~\cite{liu2024llm4cp}\\[6pt]
Signal & Received I/Q, power spectrograms, or OFDM grids & Masked signal reconstruction; contrastive learning; signal-statistic supervision & Spectrogram models~\cite{aboulfotouh2024radio,kim2026lwmspectro}, IQFM~\cite{mashaal2026iqfm}, LWM-TF (this work)\\[6pt]
Multi-input & Multiple representations, including CSI and spectrograms & Shared-backbone reconstruction or masked latent prediction & WavesFM~\cite{aboulfotouh2025wavesfm}, LatentWave~\cite{mohamed2026latentwave}\\
\hline
\end{tabular}
\end{table*}

{
\section{Link-Level Signal Generation}
\label{sec:dataset}
This section describes signal generation under the fixed link configuration
assumed in our experiments. Each sample is one
PUSCH slot of a 3GPP-compliant 5G NR uplink, simulated with
Sionna~\cite{hoydis2022sionna} along the link of Fig.~\ref{fig:rxchain}.

\begin{figure}[!t]
\centering
\includegraphics[width=\columnwidth]{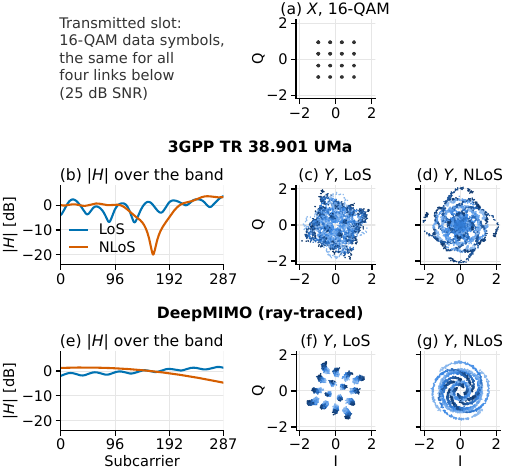}
\caption{One 16-QAM slot at $25$\,dB SNR through TR~38.901 UMa (top) and ray-traced,
timing-aligned DeepMIMO (bottom) channels, class-median links: (a) transmitted
16-QAM data symbols, common to all four links; (b), (e) $|H|$ at port~0; (c),
(d), (f), (g) received data resource elements (REs) before equalization, colored by subcarrier.}
\label{fig:channel}
\end{figure}

\subsection{Link Configuration and Transmit Chain}
\label{sec:tx}
We simulate a rank-1 SIMO CP-OFDM uplink at $f_c=3.5$\,GHz, in which a
single-antenna UE transmits one layer to a base station with two cross-polarized
($\pm45^{\circ}$) elements. Each polarization of each element is a receive port,
so the base station observes the same transmitted signal through
$N_{\mathrm{rx}}=4$ channels. Each slot has $14$ OFDM symbols, and
the UE occupies $24$ PRBs of $12$ subcarriers at $30$\,kHz spacing, that is,
$288$ subcarriers or $8.64$\,MHz. The slot length and PRB size follow the NR
numerology~\cite{3gpp38211}, whereas the allocation width and the number of
receive ports are choices of this study. The grid represents one UE's
allocation rather than a whole carrier; because each UE is demodulated on its
own PRBs and the channel statistics do not vary across the band, it models such
an allocation at any position within a wider carrier.

The transmit chain follows TS~38.211/38.212~\cite{3gpp38211,3gpp38212}
stage by stage --- transport-block CRC, code-block
segmentation, low-density parity-check (LDPC) encoding, rate matching,
scrambling, quadrature amplitude modulation (QAM) mapping, layer
mapping, and resource-grid mapping with DMRS.
{The DMRS use mapping type~A with $l_0=2$, single-symbol length, and
configuration type~1. For the full 14-symbol allocation, additional-position
settings 0--3 give the zero-based symbol sets $\{2\}$, $\{2,11\}$,
$\{2,7,11\}$, and $\{2,5,8,11\}$, respectively. With one transmitted
DMRS port, pilots occupy every second subcarrier of these symbols; the
other REs in those symbols carry no payload. See Fig.~\ref{fig:dmrs}.}

\begin{remark}[Link scope]
Every sample holds one UE on a fixed $24$-PRB SIMO rank-1 allocation,
$(N_{\mathrm{tx}},N_{\mathrm{rx}},N_{\mathrm{layers}})=(1,4,1)$, with thermal
noise as the only disturbance. A larger receive array adds antenna--PRB tokens:
the patch embedding, attention blocks, and log-power and scale targets are
unchanged, while the learned antenna--PRB position embeddings
($24N_{\mathrm{rx}}$ positions) and the relative-phase target
($2p_{\mathrm{sc}}(N_{\mathrm{rx}}-1)$ entries per patch) grow. Multi-layer transmission would also need conditioning on the precoder
and rank and per-layer targets. Variable allocations, co-scheduled users, inter-cell
interference, and transform precoding would need a variable-width or masked grid and an
interference term in~\eqref{eq:rx}; we leave them to future work.
\end{remark}

\begin{figure}[!t]
\centering
\includegraphics[width=\linewidth]{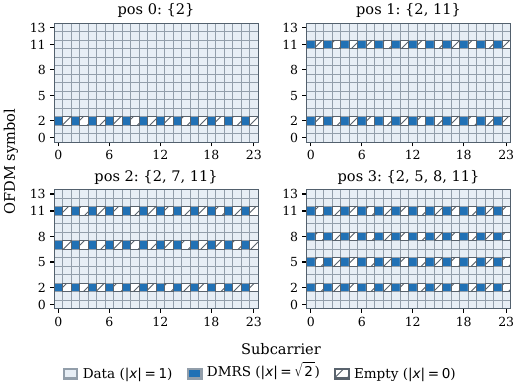}
\caption{DMRS additional positions 0--3 over two PRBs (mapping type~A, configuration
type~1): data (gray), DMRS (blue), empty (hatched).}
\label{fig:dmrs}
\end{figure}

\subsection{Channel and Noise Calibration}
\label{sec:noisecal}
For training, we use channel samples from the TR~38.901~\cite{3gpp38901}
UMa, UMi, and RMa stochastic models and the ray-traced DeepMIMO~\cite{alkhateeb2019deepmimo}
US 20 cities dataset, with DeepMIMO samples drawn uniformly across the cities.
The stochastic models include path loss and shadow fading, with one
single-sector topology drawn per sample.
Fig.~\ref{fig:channel} shows the effect on the received grid,
next to a ray-traced DeepMIMO scenario (New York,
$3.5$\,GHz)~\cite{alkhateeb2019deepmimo}.
Over $1{,}024$ UMa and $2{,}000$ ray-traced links, the median root-mean-square (RMS) delay spread is $101$/$180$\,ns (LoS/NLoS)
against $18$/$25$\,ns and the in-band $|H|$ range $10$/$21$\,dB against
$3$/$6$\,dB: UMa smears each constellation point along the band, whereas
ray-traced links mostly apply a common gain and phase ramp.

\begin{remark}[Channel model]
Stochastic scenarios keep the deployment geometry a sampled axis rather than a
fixed map, which lets the calibration below place all three scenarios on a
common SNR distribution and draws a new channel for every slot. The DeepMIMO
samples complement these stochastic channels with site-specific propagation.
\end{remark}

The channel is applied per resource element, so the received grid of a slot is
\begin{equation}
\mathbf{Y}\in\mathbb{C}^{N_{\mathrm{rx}}\times N_{\mathrm{sym}}\times
N_{\mathrm{sc}}},\qquad
Y_{a,t,f}=H_{a,t,f}\,X_{t,f}+W_{a,t,f},
\label{eq:rx}
\end{equation}
with $(N_{\mathrm{rx}},N_{\mathrm{sym}},N_{\mathrm{sc}})=(4,14,288)$, where
$X_{t,f}$ is the transmitted resource grid, $H_{a,t,f}$ the channel frequency
response at receive port $a$, OFDM symbol $t$, and subcarrier $f$, and
$W_{a,t,f}\sim\CN(0,\sigma_i^2)$ the noise of link $i$, whose variance is set
below. This is the grid a base station holds after CP removal and the FFT
(Fig.~\ref{fig:rxchain}), so the OFDM modulator and demodulator are not run.
Equation~\eqref{eq:rx} thereby neglects the inter-symbol interference of paths
beyond the cyclic prefix and the inter-carrier interference of a channel that
varies within one symbol; at $120$\,km/h the Doppler shift is at most $1.3\%$
of the subcarrier spacing.

At a fixed transmit power the three scenarios are not comparable: at $26$\,dBm
their median realized SNR spans $20.2$\,dB (UMi, $71$\,m mean range) to
$-15.4$\,dB (RMa, $1{,}749$\,m), leaving $70\%$ of RMa links below $-5$\,dB.
We therefore draw a target SNR
$\rho_{\mathrm{tgt}}\sim\mathcal{U}(-5,30)$\,dB per link, independently of every
transmitter axis, and set the noise variance from it,
\begin{equation}
\sigma_i^2=\frac{\bar{g}^{\,\gamma}\,g_i^{\,1-\gamma}}
{10^{\rho_{\mathrm{tgt},i}/10}},
\label{eq:sigma}
\end{equation}
where $g_i$ is the link's own mean channel gain and $\bar{g}$ the scenario
median, measured once per scenario from a few hundred probe links. The exponent
$\gamma\in[0,1]$ controls how much of the geometric gain spread reaches the
realized SNR: $\gamma=1$ anchors on the scenario median (the
system-level convention), while $\gamma=0$ anchors on each link (the
link-level convention of a normalized channel with a controlled SNR).

{For the linear model in \eqref{eq:rx}, increasing transmit amplitude by a factor $a>0$ with fixed noise is equivalent after per-sample RMS normalization to reducing noise amplitude by the same factor with fixed transmit amplitude: the resulting grids differ only by a common scale. A fixed-noise deployment with bounded UE power would also constrain the attainable SNR through its link budget; we do not impose such a budget in this controlled-SNR dataset.}

We use $\gamma=0$. The reason is that on UMa the LoS links
carry $43.9$\,dB more channel gain than the NLoS ones (mean over links, in
dB), so at $\gamma=1$ the realized SNR alone would identify LoS links with an
AUROC of $0.97$ --- a shortcut that any probe would exploit in place of learning
propagation structure. At $\gamma=0$ that shortcut is removed while the
structural signature survives: LoS channels remain flatter across the band
(median standard deviation of $|H|$ across the band $2.3$\,dB, against
$4.5$\,dB for NLoS, over $1{,}024$ UMa links), and the frozen encoder still
reads the LoS label at AUROC $0.82$ (Section~\ref{sec:mainresults}). With
\eqref{eq:sigma} the three scenarios share an identical SNR distribution and
differ only in their propagation structure.
{Fig.~\ref{fig:calib} shows both effects on the dataset. Within
each scenario the absolute received level still separates LoS from NLoS links
(AUROC $0.94$--$0.995$), but the per-sample RMS normalization of
Section~\ref{sec:crclabel} removes that level from the encoder input, and the
realized SNR is indistinguishable between the two (AUROC $0.50$) and across
scenarios.}

\begin{figure}[!t]
\centering
\includegraphics[width=\columnwidth]{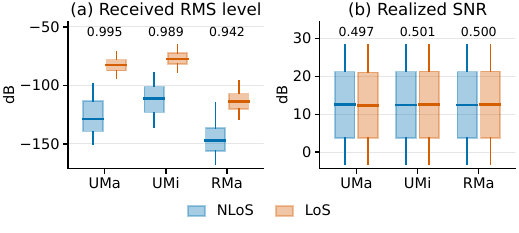}
\caption{LoS and NLoS links by scenario: (a) received RMS level, removed by the
per-sample normalization; (b) realized SNR. Boxes: quartiles; whiskers: 5th--95th
percentiles; numbers: LoS AUROC of each quantity.}
\label{fig:calib}
\end{figure}

\subsection{Link Adaptation and Modulation-Balanced Sampling}
\label{sec:sampling}
Inner-loop link adaptation (ILLA) uses the post-equalization signal-to-interference-plus-noise ratio (SINR) to propose
the highest MCS index with predicted block error rate (BLER) at most $0.1$. We apply
a static $3$\,dB SINR margin to approximate outer-loop adaptation: ILLA uses
the true channel, but the receiver decodes from a least-squares estimate. On
links drawn as in the dataset, the margin raises the CRC pass rate of the
ILLA proposal from $57\%$ to $86\%$, at a cost of $10\%$ in mean spectral
efficiency.

We then draw quadrature phase-shift keying (QPSK), 16-QAM, or 64-QAM uniformly, independently of SINR, and
map the proposal to the highest supported MCS index in that class no greater
than the proposal (or its lowest valid index if none is lower). The final MCS
fixes both modulation and code rate. We call this \emph{modulation-balanced}
sampling. In the pre-training corpus, scenario and DMRS position are cycled
in an order reshuffled every $12$ batches, and every fourth batch is
stationary. The two test sets are generated setting by setting, with every
fourth batch of each setting stationary, so each setting is $25\%$
stationary. Table~\ref{tab:paramspace} lists the parameters.

\begin{remark}[Modulation-balanced sampling]
In a deployed uplink, link adaptation uses channel quality (in particular,
SINR) to select an MCS and thus its modulation order; the receiver knows the
scheduled MCS. Modulation classification here is a representation probe.
Unconstrained ILLA would couple modulation and SNR
(rank correlation $+0.84$; QPSK/16-QAM/64-QAM frequencies
$14/20/66\%$), allowing the probe to use noise level instead of
constellation structure. To prevent this shortcut, we draw the three
modulation classes uniformly and independently of SINR
(rank correlation $-0.00$ over the $10^6$ slots; frequencies $33/33/33\%$).
The $0.1$ BLER criterion governs the ILLA proposal; the class-constrained
final MCS need not meet it.
\end{remark}

\subsection{Receiver Chain and Labels}
\label{sec:crclabel}
A 3GPP-compliant receiver runs on every sample --- least-squares (LS) channel
estimation on the DMRS, linear minimum mean-square error (LMMSE) equalization,
max-log demapping and LDPC
belief-propagation decoding with the transport-block CRC verified at the output
--- producing a per-sample decodability flag.
Every slot keeps its flag: $20.5\%$ of the $10^6$ slots fail the CRC, and
none is excluded from pre-training or evaluation.
CRC decodability collapses channel
quality, the MCS operating point and equalization tractability into one
decision-relevant binary target and is used as a downstream task in
Section~\ref{sec:downstream}.

The encoder input is $\mathbf{Y}$ itself, taken before this receiver's
channel estimation, for three reasons. First, a
channel estimate is formed from the pilot REs alone---$3.6$--$14.3\%$ of the
slot across the four DMRS settings used here (Section~\ref{sec:tx})---and
discards the data REs, which carry the modulation and any interference or
impairment. Second, the estimate depends on the estimator and interpolator a
receiver happens to use, so a representation built on it would inherit that
estimator's errors, whereas $\mathbf{Y}$ precedes every receiver-specific
choice. Third, the channel stays observable in $\mathbf{Y}$: directly on the
pilot REs, and on the data REs through their power, which follows
$|H_{a,t,f}|^2$. Equalization, one step later, would divide this channel out of
the data REs.

Each sample is stored as $\mathbf{Y}$ in half precision after per-sample RMS
normalization, which is required because the absolute receive level spans
${\sim}10^{-10}$ to ${\sim}10^{-4}$ while the smallest normal half-precision
value is $6.1\times10^{-5}$; normalizing first improves the quantization error
from $-52.9$ to $-73.7$\,dB. The discarded scale is retained as a label, so
absolute power remains available to downstream probes while the encoder sees a
numerically stable input. Alongside it we store the realized SNR
$\rho^{\mathrm{rx}}$, the target SNR, the RMS scale, the scenario, UE speed,
DMRS position, propagation distance, LoS\,/\,NLoS and indoor flags, the MCS
index and table, modulation order, code rate, transport-block size, and the CRC
and bit-error outcomes. These labels are routed exclusively to the downstream
probes of Section~\ref{sec:downstream} and are never seen by the encoder during
pre-training. The dataset used in this paper contains $10^{6}$ slots.

\begin{table}[!t]
\caption{Pre-training parameter space.}
\label{tab:paramspace}
\centering
\footnotesize
\begin{tabular}{@{}ll@{}}
\toprule
\textbf{Parameter} & \textbf{Value(s)}\\
\midrule
\multicolumn{2}{@{}l}{\emph{Fixed}}\\
\quad Carrier $f_c$            & 3.5\,GHz (FR1)\\
\quad MIMO                     & $(1,4,1)$ SIMO rank-1\\
\quad Waveform                 & CP-OFDM\\
\quad PUSCH allocation         & 24 PRB, 30\,kHz spacing (8.64\,MHz)\\
\quad Grid $\mathbf{Y}$        & $4\times14\times288$ complex\\
\addlinespace[3pt]
\multicolumn{2}{@{}l}{\emph{Sampled}}\\
\quad Scenario                 & \{UMa, UMi, RMa\}\\
\quad Target SNR               & $\mathcal{U}(-5,30)$\,dB\\
\quad Modulation               & \{QPSK, 16-QAM, 64-QAM\}, uniform\\
\quad MCS index                & SINR proposal, constrained to drawn modulation\\
\quad DMRS add.\ position      & $\{0,1,2,3\}$\\
\quad UE speed $v$             & $25\%$ at $0$, else $\mathcal{U}(0,120)$\,km/h\\
\bottomrule
\end{tabular}
\end{table}
}

\section{Model Architecture: LWM-TF}
\label{sec:arch}
\begin{figure*}[!t]
\centering
\includegraphics[width=\textwidth]{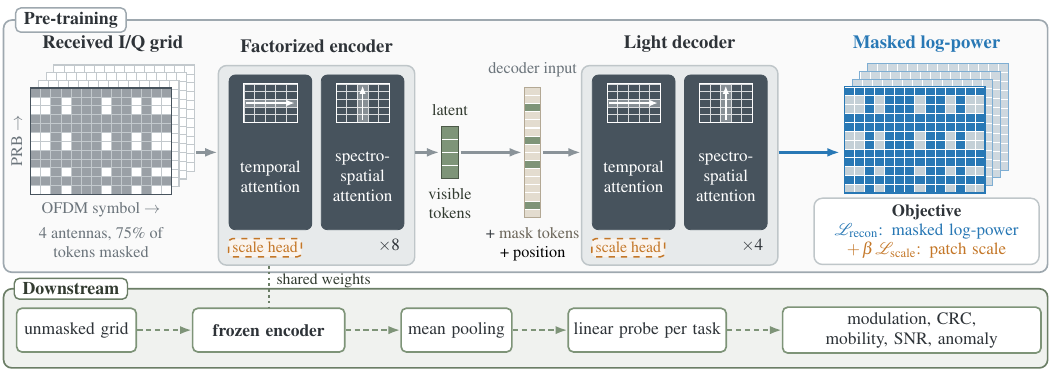}
\caption{LWM-TF overview. A factorized masked autoencoder reconstructs the log-power of
masked PRB tokens, with auxiliary patch-scale heads; downstream tasks probe the
mean-pooled frozen encoder.}
\label{fig:pipeline}
\end{figure*}
In this section, we describe the architecture of LWM-TF, a vision-style
asymmetric masked autoencoder (MAE)~\cite{he2022mae} adapted to the
receiver-side TF resource grid (Fig.~\ref{fig:pipeline}).
Section~\ref{sec:input} defines the input tokens, Section~\ref{sec:mask} the
mask schedule, Section~\ref{sec:encdec} the encoder and decoder,
Section~\ref{sec:auxheads} the auxiliary heads, and Section~\ref{sec:scaling}
the model configuration.

\subsection{Input Tensor and Patch Embedding}
\label{sec:input}
{Each input is an 8-channel real tensor of shape
$(C_{\mathrm{in}}, N_{\mathrm{sym}}, N_{\mathrm{sc}}) = (8, 14, 288)$,
containing the real and imaginary parts of the four receive signals after
per-sample RMS normalization. {Thus $8=2\times4$ counts the real
I/Q channels of the selected receive ports, $14$ spans one normal-CP slot, and
$288=24\times12$ spans only the selected PRBs.} The encoder receives neither a
pilot map nor an active-band mask. Because the allocation is fixed, it must infer the DMRS
pattern from the received grid.

A token represents one PRB on one receive antenna in one OFDM symbol, matching
the standard's allocation unit. We reshape the grid into tokens rather than
use a strided convolution, then map each patch's $2p_{\mathrm{sc}}=24$ real
I/Q values to width $d$ with one linear layer:
\begin{equation}
(n_{p,\mathrm{sym}}, n_{p,\mathrm{sp}}) =
\big(N_{\mathrm{sym}},\ N_{\mathrm{rx}}\!\cdot\!N_{\mathrm{sc}}/p_{\mathrm{sc}}\big)
= (14,\ 4\times 24),
\label{eq:tokengrid}
\end{equation}
This produces $N=14\times96=1{,}344$ tokens of width $d=256$. Separate
learned embeddings mark time and antenna--PRB position, matching the
factorized attention of Section~\ref{sec:encdec}. The I/Q input retains phase,
but no objective predicts absolute per-RE phase: the main model learns from
log-power and patch-scale targets, while P5 additionally predicts relative
phase (Section~\ref{sec:pretrain}).}

\subsection{Mask Schedule}
\label{sec:mask}
{Following the MAE recipe \cite{he2022mae}, we replace masked
patches with a learned mask token at the decoder input. The default
factorized-uniform schedule independently samples symbols and antenna--PRB
positions. It keeps $7$ of $14$ symbols and the same $48$ of $96$ spatial
positions within each, leaving $336$ of $1{,}344$ tokens visible ($75\%$
masked).}

We also compare DMRS-stratified masking, which samples pilot-bearing and data
symbols in proportion to their counts. At the same masking rate, dataset, and
compute, their three-seed means differ by at most $0.001$ on either primary task;
the seed-0 paired bootstrap intervals include zero (Fig.~\ref{fig:ablation}). We therefore use
factorized-uniform masking in the main model.

\subsection{Asymmetric Encoder--Decoder}
\label{sec:encdec}

\paragraph*{Encoder}
{The encoder is an eight-block pre-norm transformer with width
$d=256$, four attention heads, and multilayer perceptron (MLP) expansion $4$. Its attention is
factorized, as in divided space--time attention for
video~\cite{bertasius2021timesformer}: each block attends across the $14$ OFDM symbols at each
antenna--PRB position, then across the $96$ antenna--PRB positions within each
symbol, and finally applies an MLP,
\begin{align}
\mathbf{z} &\leftarrow \mathbf{z} + \MHA_{\mathrm{t}}\big(\LN(\mathbf{z})\big),\nonumber\\
\mathbf{z} &\leftarrow \mathbf{z} + \MHA_{\mathrm{s}}\big(\LN(\mathbf{z})\big),\nonumber\\
\mathbf{z} &\leftarrow \mathbf{z} + \MLPop\big(\LN(\mathbf{z})\big).
\label{eq:block}
\end{align}
This requires $\mathcal{O}\big(TS(T+S)\big)$ pair terms per layer, compared
with $\mathcal{O}\big((TS)^2\big)$ for joint attention: $148$k rather than
$1.8$M at $T=14$ and $S=96$. The encoder processes only visible tokens and
outputs features $\mathbf{H}\in\mathbb{R}^{N_{\mathrm{vis}}\times d}$ after
a final LayerNorm.}

\paragraph*{Decoder}
{The lightweight decoder has four factorized-transformer blocks,
width $d_{\mathrm{dec}}=128$, and four attention heads. It projects encoder
features from width $256$ to $128$ and inserts mask tokens to restore the full
$14\times96$ grid. After adding learned time and antenna--PRB embeddings, it
predicts each patch's log-power target (Section~\ref{sec:pretrain}). A linear
output layer maps each patch to $p_{\mathrm{sc}}=12$ values.}

\subsection{Per-Patch Auxiliary Heads}
\label{sec:auxheads}
{Three single-layer linear heads operate per patch, one on encoder
tokens and two on decoder tokens. Of the three,
only the two scale heads carry loss weight in the main model and its
predeclared ablations (Table~\ref{tab:pretraincfg}); the relative-phase head is
defined and its target validated, and only the additional P5 model of
Section~\ref{sec:ablation} trains it:}
\begin{align}
\hat{\mathbf{s}}^{\mathrm{enc}}_{i} &= \mathbf{W}_{\mathrm{enc}}\mathbf{h}_i
\in\mathbb{R}^{2},\label{eq:head_sinr}\\
\hat{\mathbf{s}}^{\mathrm{dec}}_{i} &= \mathbf{W}_{\mathrm{dec}}\mathbf{g}_i
\in\mathbb{R}^{2},\label{eq:head_phys}\\
\hat{\boldsymbol{\varphi}}_{i} &= \mathbf{W}_{\varphi}\mathbf{g}_i
\in\mathbb{R}^{6p_{\mathrm{sc}}}.\label{eq:head_amc}
\end{align}
{Here $\mathbf{h}_i$ is the $i$-th encoder token and $\mathbf{g}_i$ the
corresponding decoder token. The first two heads regress the same per-patch
scale statistic --- on the visible set from the encoder, and on the masked set
from the decoder --- at weight $0.1$ each, while the third regresses the
inter-antenna relative phase of the masked patches, at weight $0$ except in the
P5 model (Section~\ref{sec:ablation}). Precise definitions and loss terms are in
Section~\ref{sec:auxtargets}; all targets are computed deterministically from
the input alone. The heads are per patch rather than per sample so
that the scale statistic, which patch normalization removes from
$\mathcal{L}_{\mathrm{recon}}$, is restored at the same granularity at which
it was removed.}

\subsection{Model Configuration and Scaling}
\label{sec:scaling}
Table~\ref{tab:modelcfg} lists the base configuration. {Total parameter
count is 9.58\,M, dominated by the encoder (8.46\,M, of which 8.43\,M is the eight
factorized blocks); the decoder accounts for 1.12\,M, comprising the four
128-wide blocks (1.06\,M), the encoder-to-decoder projection, the reconstruction,
scale and relative-phase heads, the mask token, and the position and layer-norm
parameters.} Each model is pre-trained for 5,000 optimizer steps, and the main
model is then continued to 40,000 steps (Section~\ref{sec:traincfg}).
The dataset size, training pools, sample presentations, and token counts are
reported separately in Table~\ref{tab:pretraincfg} and
Section~\ref{sec:traincfg}.

\begin{table}[!t]
\caption{Model configuration.}
\label{tab:modelcfg}
\centering
\footnotesize
\setlength{\tabcolsep}{4pt}
\begin{tabular}{@{}ll@{}}
\toprule
\textbf{Component} & \textbf{Value}\\
\midrule
\multicolumn{2}{@{}l}{\emph{Patch embedding}}\\
\quad Patch $(p_{\mathrm{sym}}, p_{\mathrm{sc}})$ & $(1, 12)$: one PRB, symbol, antenna\\
\quad Input features per patch & $24$ ($12$ subcarriers, I and Q)\\
\quad Tokens per sample $N$ & $1{,}344 = 14 \times (4\times 24)$\\
\quad Visible tokens at $75\%$ masking & $336$\\
\quad Position embedding & learned: $14$ temporal, $96$ spatial\\
\addlinespace[3pt]
\multicolumn{2}{@{}l}{\emph{Encoder} (temporal, then spectro-spatial)}\\
\quad Depth\,/\,hidden\,/\,heads & 8\,/\,256\,/\,4\\
\quad MLP expansion & $4\times$\\
\addlinespace[3pt]
\multicolumn{2}{@{}l}{\emph{Decoder (asymmetric)}}\\
\quad Encoder-to-decoder projection & $\mathbb{R}^{256}\rightarrow\mathbb{R}^{128}$\\
\quad Depth\,/\,hidden\,/\,heads & 4\,/\,128\,/\,4\\
\quad Reconstruction head & $\mathbb{R}^{128}\rightarrow\mathbb{R}^{12}$ (log-power)\\
\addlinespace[3pt]
\multicolumn{2}{@{}l}{\emph{Auxiliary heads (per patch)}}\\
\quad Encoder scale head & $\mathbb{R}^{256}\rightarrow\mathbb{R}^{2}$, weight $0.1$\\
\quad Decoder scale head & $\mathbb{R}^{128}\rightarrow\mathbb{R}^{2}$, weight $0.1$\\
\quad Relative-phase head & $\mathbb{R}^{128}\rightarrow\mathbb{R}^{72}$, weight $0$ ($1$ in P5)\\
\addlinespace[3pt]
\multicolumn{2}{@{}l}{\emph{Mask schedule}}\\
\quad Default & factorized uniform, $75\%$ masked\\
\quad Ablation & DMRS-stratified, $75\%$ masked\\
\addlinespace[3pt]
\textbf{Total parameters} & \textbf{9.58\,M}\\
\quad Encoder & 8.46\,M\\
\quad Decoder (incl.\ projection and heads) & 1.12\,M\\
\bottomrule
\end{tabular}
\end{table}

\section{Pre-Training: Masked Log-Power Reconstruction with Per-Patch Scale
Targets}
\label{sec:pretrain}
In this section, we describe the label-free pre-training objective.
Section~\ref{sec:multiloss} combines its terms, Section~\ref{sec:recon}
defines the masked log-power reconstruction, Section~\ref{sec:auxtargets} the
auxiliary targets, and Section~\ref{sec:traincfg} the training configuration.

\subsection{Multi-Loss Objective}
\label{sec:multiloss}
LWM-TF is trained against a fully label-free objective that combines a masked
{log-power reconstruction with two per-patch scale terms and an
optional relative-phase term:}
\begin{equation}
\mathcal{L} = \mathcal{L}_{\mathrm{recon}}
+ \beta_{\varphi}\,\mathcal{L}_{\varphi}
+ \beta_{\mathrm{enc}}\,\mathcal{L}^{\mathrm{enc}}_{\mathrm{scale}}
+ \beta_{\mathrm{dec}}\,\mathcal{L}^{\mathrm{dec}}_{\mathrm{scale}},
\label{eq:total_loss}
\end{equation}
{with $\beta_{\mathrm{enc}}=\beta_{\mathrm{dec}}=0.1$ and
$\beta_{\varphi}=0$ in the main model and its predeclared ablations
($\beta_{\varphi}=1$ only in the P5 model of Section~\ref{sec:ablation}).
All targets are computed
deterministically from the input TF grid; no downstream label, pilot layout,
or simulator quantity is consumed.}

\subsection{Masked Log-Power Reconstruction}
\label{sec:recon}
Let $\mathcal{M}\subset\{1,\dots,N\}$ be the index set of patches replaced by
the mask token (Section~\ref{sec:mask}), and let
$\mathbf{p}_p\in\mathbb{R}^{p_{\mathrm{sym}}p_{\mathrm{sc}}}$ denote the
per-patch log-power
\begin{equation}
\mathbf{p}_p = \log\!\left(\max\{|\mathbf{y}|^2,10^{-12}\}\right)
\Big|_{\text{patch }p}.
\label{eq:logpower}
\end{equation}
{The reconstruction target is this log-power normalized by its own
patch statistics, $\tilde{\mathbf{p}}_p=(\mathbf{p}_p-\mu_p\mathbf{1})/
\sqrt{v_p+\epsilon}$, where $\mu_p$ and $v_p$ are the mean and variance of the
$12$ entries of $\mathbf{p}_p$ (the same statistics that form the scale target
\eqref{eq:ssinr}) and $\epsilon=10^{-6}$. The decoder produces
$\hat{\mathbf{p}}_p$ for every patch, and the reconstruction loss is the mean
squared error over the masked patches,
\begin{equation}
\mathcal{L}_{\mathrm{recon}} =
\frac{1}{|\mathcal{M}|}\sum_{p\in\mathcal{M}}
\frac{1}{p_{\mathrm{sc}}}\big\|\hat{\mathbf{p}}_p - \tilde{\mathbf{p}}_p\big\|_2^2 .
\label{eq:lrecon}
\end{equation}
Every RE of a masked patch contributes, including DMRS and empty pilot-comb
REs; the input spans only the $288$ allocated subcarriers and so contains no
guard band (Section~\ref{sec:tx}). Normalization keeps
high-power patches from dominating the loss and removes the shortcut of
predicting a patch's level, which the scale target supervises separately.}

\begin{figure}[!t]
\centering
\includegraphics[width=\columnwidth]{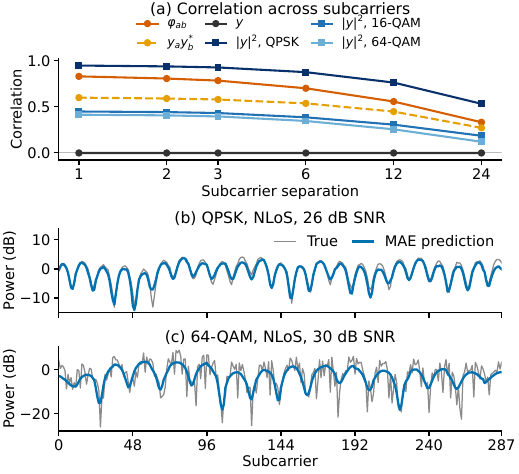}
\caption{(a) Correlation across subcarriers at separation $\Delta$ of $y$,
$y_a y_b^{*}$, its relative phase $\varphi_{ab}$, and $|y|^2$ by modulation (median over $900$
held-out slots at $18$--$22$\,dB, data symbols; pooled over modulations, $|y|^2$
gives $0.49$ at $\Delta=1$). (b), (c) Masked log-power
reconstruction by P0 on an OFDM symbol hidden from the encoder, rescaled with
the patch mean and variance that P0 predicts (antenna~0; median-error NLoS
slot, SNR $\geq 20$\,dB).}
\label{fig:recon}
\end{figure}

{
\paragraph[Why is the target power rather than the complex grid?]{Why is the target power rather than the complex grid?\noheadpunct}
Reconstructing $\mathbf{y}$ itself is the obvious alternative. The transmit chain scrambles the codeword before modulation
mapping, and on held-out slots at $18$--$22$\,dB we observe an
adjacent-subcarrier correlation of $0.00$ for $y$ against $0.49$ for $|y|^2$
(Fig.~\ref{fig:recon}(a)). We therefore model the payload symbols as
independent and zero-mean. Under this approximation, for a masked data
resource element
\begin{equation}
\mathbb{E}\big[\,y \,\big|\, \text{context}\,\big]
= \mathbb{E}\big[H \,\big|\, \text{context}\big]\, \mathbb{E}[X] = 0 ,
\label{eq:degenerate}
\end{equation}
so under this approximation the $L_2$-optimal prediction of such an element
is zero, and a complex reconstruction target carries no conditional-mean
signal on it. Under
the factorized mask of Section~\ref{sec:mask}, this applies to the two thirds
of masked tokens that lie in fully masked OFDM symbols. A masked token in a
kept symbol can share its payload symbols with a visible token of another
antenna on the same PRB, and its conditional mean is then not zero. The
complex target thus has a weak signal, or is \emph{degenerate}, on most, but
not all, masked data REs, which is consistent with the complex-reconstruction model still learning
a useful representation (Section~\ref{sec:ablation}).

This is a consequence of the input interface rather than of the objective.
Models supervised on the channel $\mathbf{H}$ never meet it, because
$\mathbf{H}$ is predictable from its neighbours; it appears only once the
encoder is handed the received signal. Power is well posed where phase is
not. Under the model above, $y$ has zero conditional mean, whereas the
channel-power proxy $\log(|H|^2+\sigma^2)$ has a squared correlation of $0.45$
with $\log|y|^2$ over the data REs of RMS-normalized grids ($0.62$ at
$18$--$22$\,dB; $768$ static-UE links in UMa, UMi, and RMa, SNR from
$\mathcal{U}(-5,30)$\,dB). Patch normalization moves each token's mean and variance to
the scale target, and the token means of the two have a squared correlation
of $0.87$; the normalized target keeps the within-token pattern, whose squared
correlation with the likewise normalized proxy is $0.09$ ($0.19$ for QPSK),
the rest including payload amplitudes, noise, and channel structure that the
proxy misses. What \eqref{eq:degenerate} weakens is the \emph{per-element}
complex target;
Section~\ref{sec:auxtargets} shows that a second-order statistic recovers the
phase that \eqref{eq:logpower} discards without reintroducing the degeneracy.
}

\paragraph{Predictable structure and the per-RE error floor}
At a payload RE, log-power reflects channel power, symbol power, and noise:
\begin{equation}
\log|y[k,\ell]|^2 \approx
\underbrace{\log|h[k,\ell]|^2}_{\text{smooth in }(k,\ell)}
+ \underbrace{\log|x[k,\ell]|^2}_{\text{i.i.d.\ across REs}}
+ \ \text{noise},
\label{eq:decomp}
\end{equation}
The channel term varies smoothly across subcarriers and symbols, so nearby
REs help predict it. Payload symbols and receiver noise are random. In our
mask, two thirds of masked tokens lie in fully hidden OFDM symbols, so the
encoder sees no antenna carrying those symbols. For some masked tokens in
visible symbols, another antenna may show the same transmitted symbol. The
model can therefore learn the shared channel envelope, but not every data-RE
value exactly; random symbol amplitudes and noise leave an error floor.

\begin{figure}[!t]
\centering
\includegraphics[width=\columnwidth]{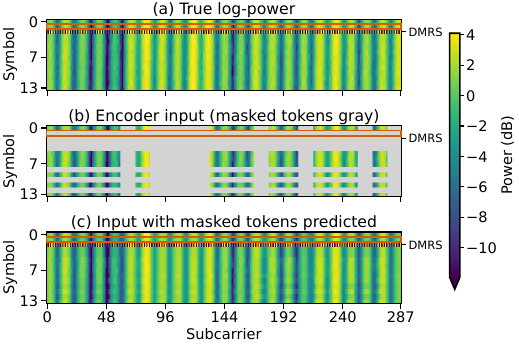}
\caption{One masked slot (QPSK, NLoS, $26$\,dB; antenna~0): (a) true
log-power; (b) encoder input, masked tokens gray; (c) input with masked tokens
predicted by P0 and rescaled with the patch mean and variance that P0
predicts. The orange box marks symbol~$1$, the hidden data symbol
plotted in Fig.~\ref{fig:recon}(b). The striped row below it is the DMRS
symbol~$2$ (Fig.~\ref{fig:dmrs}): its empty comb REs carry only noise and fall
below the color scale. It is hidden in (b) and restored in (c).}
\label{fig:maskgrid}
\end{figure}

{Patch normalization gives each target zero mean and unit variance,
so predicting zero yields a mean squared error (MSE) of exactly $1$. When we hide a whole OFDM
symbol, the decoder follows the channel envelope, reaching $R^2=0.62$ per RE
and $0.85$ for the patch mean over $256$ test slots. Fig.~\ref{fig:maskgrid}
shows one masked slot in full: the factorized mask hides $7$ of the $14$
symbols and, in the others, the same antenna--PRB positions, and the
prediction restores the fading pattern in both kinds of hidden region. For QPSK, whose symbol
amplitude is fixed, the envelope explains nearly all the signal variation;
with 64-QAM, random symbol amplitudes remain. Reconstruction loss is therefore
a pre-training signal, not a measure of representation quality; we assess the
encoder on downstream tasks (Section~\ref{sec:downstream}). The scale heads
restore the patch level removed by normalization, while relative phase remains
unsupervised in the main model (Sections~\ref{sec:auxtargets}
and~\ref{sec:ablation}).}

{
\subsection{Auxiliary Targets}
\label{sec:auxtargets}
The main configuration uses patch-scale supervision in addition to
reconstruction. Inter-antenna relative phase is an optional auxiliary
quantity; its loss weight is zero in the main configuration. These
quantities are computed from the input grid without downstream labels.

\paragraph{Patch scale ($\mathcal{L}_{\mathrm{scale}}$)}
Patch normalization removes the per-patch mean and variance. We retain
this information through the following auxiliary target.
For each receive antenna, a patch contains 12 consecutive subcarriers
within one OFDM symbol. Let $y_{i,q}$ be the complex, per-slot
RMS-normalized received sample at subcarrier $q$ of patch $i$. We compute
\begin{align}
p_{i,q} &= \log\!\left(\max\{|y_{i,q}|^2,10^{-12}\}\right),
\qquad q=1,\ldots,12, \nonumber\\
\mu_i &= \frac{1}{12}\sum_{q=1}^{12}p_{i,q}, \qquad
v_i = \frac{1}{12}\sum_{q=1}^{12}(p_{i,q}-\mu_i)^2, \nonumber\\
\mathbf{s}_i &= [\mu_i,\ \log(v_i+10^{-6})]^{\top}.
\label{eq:ssinr}
\end{align}
All logarithms are natural, and the variance uses divisor 12. These statistics use all received REs within
the patch, including non-pilot REs in a DMRS-bearing symbol; no pilot
interpolation or averaging across DMRS symbols is performed. Thus, the
target remains defined with only one DMRS-bearing symbol and even with
zero within-patch variance. It describes received log-power, not an
estimate of SINR or channel power.

A check of all $10^6$ samples, covering all four DMRS
settings in all three scenarios, found no non-finite received samples,
log-power targets, or relative-phase targets. No received power fell below
$10^{-12}$, and no patch variance fell below $10^{-6}$. The scale targets
are therefore finite throughout the dataset, and no samples were discarded
because of invalid targets. The power floor is a numerical safeguard;
the $10^{-6}$ variance offset is applied to every patch.

The scale target is regressed twice: from the encoder tokens on the visible set $\mathcal{K}$, so
that the level survives into the latent a probe reads, and from the decoder
tokens on the masked set $\mathcal{M}$, so that the reconstruction path
recovers it as well,
\begin{align}
\mathcal{L}^{\mathrm{enc}}_{\mathrm{scale}} &=
\frac{1}{2|\mathcal{K}|}\sum_{i\in\mathcal{K}}
\big\|\hat{\mathbf{s}}^{\mathrm{enc}}_i - \mathbf{s}_i\big\|_2^2,
\label{eq:sphys}\\[2pt]
\mathcal{L}^{\mathrm{dec}}_{\mathrm{scale}} &=
\frac{1}{2|\mathcal{M}|}\sum_{i\in\mathcal{M}}
\big\|\hat{\mathbf{s}}^{\mathrm{dec}}_i - \mathbf{s}_i\big\|_2^2 .
\label{eq:sphys2}
\end{align}

\paragraph{Inter-antenna relative phase ($\mathcal{L}_{\varphi}$)}
Reconstructing log-power leaves phase out of the training signal altogether,
and by \eqref{eq:degenerate} reconstructing $\mathbf{y}$ would restore it only
weakly. That weakness, however, concerns a \emph{first} moment. The second moment does not share it, and that is what we use.

Collect the receive ports at one resource element into
$\mathbf{y}[k,\ell]\in\mathbb{C}^{N_{\mathrm{rx}}}$ and form the per-RE spatial
covariance $\mathbf{R} = \mathbf{y}\mathbf{y}^{\mathsf{H}}$, whose entries are
\begin{equation}
[\mathbf{R}]_{aa} = |y_a|^2 ,
\qquad
[\mathbf{R}]_{ab} = y_a y_b^{*},\quad a\neq b .
\label{eq:cov}
\end{equation}
Its diagonal is exactly the quantity \eqref{eq:logpower} already reconstructs.
Its off-diagonal is new; under the same model, with noise independent across
ports and unit symbol power,
\begin{equation}
\mathbb{E}\big[\,y_a y_b^{*}\,\big|\,\mathbf{H}\,\big]
= H_a H_b^{*}\;\mathbb{E}\big[|X|^2\big]
= H_a H_b^{*} ,
\label{eq:secondmoment}
\end{equation}
because the payload enters through $|X|^2$ rather than $X$. Set
\eqref{eq:secondmoment} against \eqref{eq:degenerate}: the same unpredictable
symbol that forces the first moment to vanish leaves the second moment
proportional to the channel. \emph{The current target is the diagonal of the
right object, and the off-diagonal is a phase-bearing target that is free of
this degeneracy.}

For a single resource element $|y_a y_b^{*}| = |y_a|\,|y_b|$ exactly, so the
magnitude of \eqref{eq:cov} is already carried by the diagonal. Dividing it
out leaves the relative phase alone,
\begin{equation}
\varphi_{ab} = \frac{y_a y_b^{*}}{\big|y_a y_b^{*}\big|}
= e^{\,\jmath(\angle y_a - \angle y_b)} \in \mathbb{C},
\qquad |\varphi_{ab}| = 1 ,
\label{eq:samc}
\end{equation}
which the decoder regresses as a $(\cos,\sin)$ pair --- a parameterization that
has no wrap-around --- over the $N_{\mathrm{rx}}-1$ partner ports of each
patch,
\begin{equation}
\mathcal{L}_{\varphi} = \frac{1}{|\mathcal{M}|}\sum_{i\in\mathcal{M}}
\frac{1}{2p_{\mathrm{sc}}(N_{\mathrm{rx}}-1)}
\big\|\hat{\boldsymbol{\varphi}}_i - \boldsymbol{\varphi}_i\big\|_2^2 .
\label{eq:lphi}
\end{equation}
Predicting zero scores $\tfrac{1}{2}$ on \eqref{eq:lphi} since
$|\varphi_{ab}|=1$, which fixes the scale against which convergence is read.
At $\beta_{\varphi}=1$, the component-averaged relative-phase and
reconstruction terms of \eqref{eq:total_loss} have equal weight.

Three measurements support the construction. Against the true
$H_a H_b^{*}$, \eqref{eq:samc} tracks the channel phase with a median error of
$5.3^{\circ}$ at $18$--$22$\,dB (complex correlation $0.96$ of the
unit-modulus terms) and $12^{\circ}$ over the full SNR range. The normalization is what makes the target
informative rather than redundant: at $18$--$22$\,dB the adjacent-subcarrier
correlation of $\varphi_{ab}$ is $0.83$ against $0.60$ for the raw product
$y_a y_b^{*}$, which is dominated by the magnitude already on the diagonal
(Fig.~\ref{fig:recon}(a)). And the degeneracy does not return: the same
correlation is $0.00$ for $y$ itself.

Finally, the absolute phase $\angle y_a$ of one resource element carries the
payload symbol, which we model as unpredictable; only \emph{relative} phase
--- across ports, and across symbols --- is learnable. Because
$\varphi_{ab}$ compares two ports at the same resource element, any phase
rotation common to all ports cancels in it, so the target follows the spatial
structure of the channel.

\paragraph[Why are the targets per patch?]{Why are the targets per patch?\noheadpunct}
Both auxiliary losses are computed per patch rather than on the mean-pooled
readout. The quantity each restores was removed per patch --- the scale by
patch normalization, the phase by reconstructing power --- so supervising at
the same granularity is what makes the restoration exact. An earlier design
that attached per-sample statistics to the mean-pool readout was simpler but
could only recover a single scalar per slot, which is not enough to describe a
frequency-selective channel.
}

\begin{table}[!t]
\caption{Pre-training configuration and reported continuations
(cumulative counts per run).}
\label{tab:pretraincfg}
\centering
\footnotesize
\begin{tabular}{@{}p{0.43\linewidth}p{0.51\linewidth}@{}}
\hline
\rule{0pt}{8.2pt}\textbf{Item} & \textbf{Configuration}\\
\hline
Dataset size & 1,000,000 slots\\
P0--P5 training pool & 850,496 slots\\
Development/validation partition & 149,504 slots\\
Source-only D0 training pool & 423,936 slots (UMa/UMi, DMRS 0--2)\\
Target SNR, all runs & $\mathcal{U}(-5,30)$\,dB per link\\
Noise calibration & Per-link channel gain ($\gamma=0$)\\
Training seeds & 0, 2, 3; beyond 40k: P0 seed 0 only\\
Effective batch size & 512\\
Reported total steps & 5k: P0--P5, D0; 40k: P0, P5, D0; 60k/80k: P0 seed 0\\
Presentations at 5k/40k/80k & 2.56\,M / 20.48\,M / 40.96\,M\\
Tokens per sample (full/visible) & 1,344 / 336\\
Full-grid positions at 5k/40k/80k & $3.44\times10^{9}$ / $2.75\times10^{10}$ / $5.51\times10^{10}$\\
Visible positions at 5k/40k/80k & $8.60\times10^{8}$ / $6.88\times10^{9}$ / $1.38\times10^{10}$\\
Optimizer & AdamW ($\beta_1=0.9$, $\beta_2=0.95$)\\
Peak learning rate & $3\times10^{-4}$ initially; $10^{-4}$ in continuations\\
Initial schedule & 1,000-step warmup, cosine to step 5,000\\
Continuation schedules & 500-step warmup, cosine to 40k; P0 seed 0 restarts the cycle toward 100k (details in text)\\
Weight decay & 0.05\\
Gradient clipping ($\ell_2$) & 1.0\\
Precision & bfloat16 mixed precision\\
Default masking & Factorized uniform; 75\% masked\\
Masking ablation & DMRS-stratified; 75\% masked\\
P0/D0 loss weights & Reconstruction 1; encoder/decoder scale 0.1 each; phase 0\\
P5 loss weights & P0 weights plus phase 1\\
Evaluation checkpoints & 5k/40k endpoints; 10k/20k and seed-0 60k/80k horizon diagnostics\\
\hline
\end{tabular}
\end{table}

\subsection{Training Configuration}
\label{sec:traincfg}
\begin{figure}[!t]
\centering
\includegraphics[width=\columnwidth]{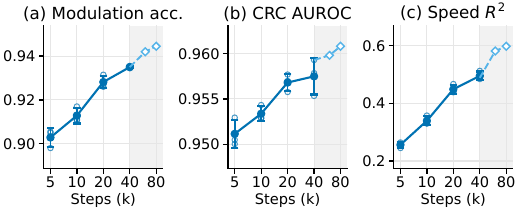}
\caption{Development-validation scores of the frozen main model after
$5{,}000$ pre-training steps and at $10{,}000$, $20{,}000$, and $40{,}000$ steps
of the continuation (linear probe fitted on $8{,}000$ labels and scored on
$2{,}000$ development-validation labels; mean $\pm$ standard deviation (SD) and individual seeds 0,
2, and 3). Diamonds (shaded): seed~0 continued from $40{,}000$ steps
in a second warm-up and cosine cycle toward $100{,}000$. No test data are
used.}
\label{fig:horizon}
\end{figure}

{
The $10^6$-slot dataset contains 850,496 pre-training and 149,504
development/validation slots. P0 combines log-power reconstruction with scale
supervision; P1--P4 respectively remove scale supervision, use scale alone,
reconstruct complex I/Q, or use DMRS-stratified masking. D0 uses P0's
objective on 423,936 source-only slots (UMa/UMi, DMRS positions 0--2).
All runs keep each link's stored target SNR, uniform on $[-5,30]$\,dB,
with per-link noise calibration ($\gamma=0$).

P0--P4 and D0 initially use 5,000 steps, batch size 512, and seeds 0, 2,
and 3. This is the matched budget of the ablations and Table~VII.
The main P0 model continues all three runs to 40,000 total steps, with a new
optimizer, 500-step warm-up to $10^{-4}$, and cosine decay to zero.
The horizon was chosen on seed-0 development validation only, by mean
modulation accuracy and CRC AUROC (labels: Fig.~\ref{fig:horizon}).
The additional D0 and P5 runs use the same continuation to 40,000 steps
with three seeds. D0 retains its source pool; P5 adds relative-phase loss
at weight 1 to P0 on the full pre-training pool.

P0 seed~0 then retains its optimizer state and starts a second 500-step
warm-up to $10^{-4}$ at 40,000, with cosine decay toward 100,000.
Its 60,000/80,000 checkpoints are exploratory mid-cycle diagnostics;
the main results use the three-seed 40,000-step model. The 5,000/40,000
comparisons use schedule endpoints.

Table~IV reports cumulative exposure: $512n$ presentations after $n$ total
steps, with 1,344 full-grid and 336 visible token positions per presentation.
Thus 60,000 steps gives 30.72\,M presentations; the table lists the
5,000/40,000/80,000 totals. These count repeated visits, not distinct slots
or processing across transformer layers.
Table~\ref{tab:pretraincfg} summarizes the initial runs and the reported
continuations, with cumulative counts per run. Supervised-baseline and
label-budget experiments do not add encoder pre-training steps.
}
Fig.~\ref{fig:horizon} shows the development-validation scores
behind the $40{,}000$-step horizon, the last checkpoint all three
runs share; all increase up to $40{,}000$ steps. Continued to $80{,}000$ steps,
seed~0 gains in modulation accuracy ($0.935$ to $0.945$) and speed $R^2$
($0.49$ to $0.60$) but not in CRC AUROC, so the $40{,}000$-step results are
conservative.

\begin{remark}[Pre-training horizon]
The continuation improves every task: on test set~1
(Section~\ref{sec:protocol}), modulation accuracy rises from $0.906$ to
$0.938$, CRC AUROC from $0.953$ to $0.962$, speed $R^2$ from $0.275$ to
$0.529$, and mobility AUROC with $1{,}000$ labels from $0.75$ to $0.89$. The
representation is thus not saturated at $5{,}000$ steps.
\end{remark}

\section{Data Representation}
\label{sec:datarep}
In this section, we describe the representation that the downstream probes
receive. The encoder takes the normalized $8\times14\times288$ real I/Q tensor of
Section~\ref{sec:input}, formed from the four-antenna complex received grid
\eqref{eq:rx}. Phase is present in the input, although the main pre-training
targets supervise only log-power and per-patch scale.

At inference the encoder processes the full, unmasked grid ($N=1{,}344$
tokens, four times the pre-training visible count). Every downstream probe
receives only the mean of its final LayerNorm token features,
\begin{equation}
\bar{\mathbf{e}} = \frac{1}{N}\sum_{p=1}^{N}\mathbf{H}_p\in\mathbb{R}^{256},
\label{eq:meanpool}
\end{equation}
with no decoder output, per-token features, or metadata. CRC and other
physical labels are stored separately and never enter pre-training; its
targets are deterministic functions of the input grid. The DMRS-stratified
masking ablation uses DMRS metadata only to place masks.

Unlike a short-time Fourier transform spectrogram, this grid holds complex, standard-aligned
NR resource elements. This alignment defines the PRB tokens and lets the
receiver-statistics control measure noise on empty DMRS-comb REs.

\section{Downstream Evaluation}
\label{sec:downstream}
\subsection{Tasks}
\label{sec:tasks}
Each task is read out by a linear probe on the pooled embedding
$\bar{\mathbf{e}}$ of Section~\ref{sec:datarep}, with the pre-trained encoder
of Section~\ref{sec:pretrain} kept frozen. Two tasks, modulation
classification and CRC decodability, were fixed as primary before the final
scoring; channel time-variation detection was added after the primary tasks
were fixed, under its own frozen protocol. SNR estimation serves as a control, and held-out waveform
impairments as a stress test.

{\paragraph{Modulation classification (3 classes, accuracy)}
The target is the modulation order (QPSK, 16-QAM, or 64-QAM). Under
modulation-balanced sampling (Section~\ref{sec:sampling}) the classes are
equally frequent and uncorrelated with SNR, so the label cannot be read from
the noise level; recovering it requires the constellation structure of the
data REs.}

\paragraph{CRC decodability (binary, AUROC)}
The frequency-domain LDPC\,+\,LMMSE\,+\,LS receiver of
Section~\ref{sec:crclabel} produces a per-sample CRC pass\,/\,fail flag, used
here as the binary target \cite{strodthoff2019harq}. CRC decodability collapses
channel quality, modulation order, code rate, and equalization tractability
into a single decision-relevant scalar. {The label is specific to
this receiver; a probe predicts whether it will decode, it does not decode.}
Predicting the CRC outcome before the iterative LDPC decoder runs has been
studied as early hybrid automatic repeat request (HARQ) feedback prediction in 5G
\cite{strodthoff2019harq}; a frozen-encoder predictor on the post-FFT grid
alone, without demodulated symbols or soft log-likelihood ratios, addresses the same decision.

{\paragraph{Channel time-variation (mobility) detection (binary, AUROC)}
Links with UE speed $v\ge20$\,km/h are separated from static and slower links.
The threshold is where the receiver should switch its DMRS interpolation from
time averaging to linear interpolation: at $15$\,dB the normalized estimation mean squared error is
$-23.0$ against $-21.0$\,dB for a static UE and $-11.9$ against $-20.4$\,dB at
$120$\,km/h. The label is the speed, but what the grid carries is the channel's
variation over the slot, $0.2$ to $1.2$\,rad of Doppler phase at $20$ to
$120$\,km/h, and a base station has no physical-layer speed report to read it
from. The task was added after the primary tasks were fixed, with its methods, budgets, and
contrasts fixed before it was scored on either test set.

\paragraph{SNR estimation (regression, $R^2$; control)}
The target is the realized per-slot SNR $\rho^{\mathrm{rx}}$ of
Section~\ref{sec:crclabel}. On this input a parameter-free receiver estimator
recovers it from the empty REs of the DMRS comb (Section~\ref{sec:protocol}),
so the task tests whether a representation preserves a quantity
already available in closed form, not whether it adds information.

\paragraph{Waveform anomaly detection (binary, AUROC)}
Clean slots are separated from slots carrying one of three waveform-domain
impairments---narrowband interference (NBI), an impulsive burst, or receiver
saturation followed by coarse quantization---injected after automatic gain
control and before CP removal and the FFT. The detectors see clean grids
only, so every family is unseen (Section~\ref{sec:anomaly},
Appendix~\ref{app:anomaly}).}

{Five further labels---DMRS position, LoS\,/\,NLoS, scenario,
distance, and UE speed as a regression target---are read out by the same
frozen probes as secondary diagnostics (Section~\ref{sec:mainresults}).}

\subsection{Probe and Evaluation Protocol}
\label{sec:protocol}

{
\paragraph{Data splits}
The dataset is partitioned into 850,496 samples for pre-training and
149,504 for development/validation. A group is a contiguous run with the
same scenario, DMRS configuration, and user speed, not a UE identifier:
the simulator draws a new topology for each slot. Each group stays in
one partition.

For the 10,000-label experiments, 8,000 training samples are drawn from
the pre-training set and 2,000 validation samples from the validation set.
Performance is evaluated on a separately generated test set of 49,152
samples (test set~1), with 4,096 samples for each scenario--DMRS combination. Test
samples are not used for pre-training, downstream training, or checkpoint
selection. Follow-up experiments designed after inspecting the test results
use the same test set and are reported separately from the core comparison.

\paragraph{Scope of generalization}
The main evaluation uses new samples from the same scenarios and simulator
as the training data. To evaluate performance under unseen conditions,
we also train a model using only UMa and UMi scenarios with DMRS
additional positions 0--2. Both pre-training and downstream training and
validation use only these conditions. We then evaluate on RMa, DMRS
additional position~3, and their combination, alongside the training
conditions. This evaluates generalization to unseen scenarios
and DMRS configurations within the TR~38.901 simulator.

\paragraph{Two test sets}
The core results on the $49{,}152$-slot test set (test set~1) were examined
before the added controls and low-label comparisons were designed.
Test set~2 ($24{,}576$ slots, $2{,}048$ per scenario--DMRS cell in
$192$ generation batches) was generated with a different random seed after
the original $64$ evaluation candidates (checkpoints, probe heads, and
supervised fits) were hash-frozen; those candidates were scored without
modification. The $40{,}000$-step main model, whose horizon was chosen on
seed-0 development validation, and the seed-0 continuation toward
$100{,}000$ steps were added afterwards and scored with the same probe
protocol; test set~2 selects neither the pre-training horizon nor the probes.
We report the two test sets separately and never pool them.

\paragraph{Frozen-encoder probes}
The embedding $\bar{\mathbf{e}}$ is standardized with fit-split statistics and
augmented by a bias. Classification uses multinomial logistic regression with
an $\ell_2$ penalty of $5\times10^{-4}$, fitted by L-BFGS until the gradient
norm falls below $10^{-5}$; regression uses ridge regression with unit
penalty. No probe hyperparameter is tuned, and the validation labels are
unused by the frozen probes.

\paragraph{Controls}
(i)~\emph{Random-init encoder}: the same architecture with random weights and
the same probe. (ii)~\emph{Receiver statistics}: seven statistics computed from
the grid and the fixed receiver configuration (Appendix~\ref{app:aux}),
read out by the same linear probe and by a $7\!\rightarrow\!64\!\rightarrow\!64$
GELU MLP. The excess of data-RE over empty-RE power, relative to the latter,
gives a \emph{parameter-free} SNR estimate. (iii)~\emph{Supervised CNN}: a
$55.6$\,k-parameter network trained end to end on the grid; its input is the
I/Q of the four antennas plus $\log(1+|y|^2)$, followed by two pointwise
($32$ channels) and two $3\times3$ convolutions ($48$, $64$; $1\times4$
subcarrier pooling between), mean and variance pooling of both stages,
and a $192\!\rightarrow\!64$ MLP head (dropout $0.3$). (iv)~\emph{Scratch transformer}: the LWM-TF encoder with a linear head,
trained end to end from random initialization. The modulation/CRC controls
use AdamW (weight decay $0.01$), a plateau scheduler, and patience $20$.
Seed~0 selects the learning rate by the budget's validation loss from
$\{10^{-4},3\!\times\!10^{-4},10^{-3}\}$ (scratch transformer:
$\{10^{-5},3\!\times\!10^{-5},10^{-4}\}$); seeds 2 and 3 reuse that rate
within each task and budget. Full-budget CNN runs were repeated from
initialization with a $400$-epoch cap instead of $100$; all six stopped by
patience. Low-budget CNNs use a $200$-epoch cap and batch $256$ without
augmentation; the MLP cap is $300$. We report the extended CNN and retain
the original full-budget runs as a sensitivity result.
(v)~\emph{Doppler features} (mobility only): the same MLP on
the seven statistics plus $26$ pilot-free statistics, a task-specific control
built from receiver knowledge: per antenna pair, the relative phase
$u=y_a y_b^{*}/|y_a y_b|$, which cancels the payload, gives $|\Mean_t u|$ and
the magnitudes of its lag-$1$ correlation across symbols and lag-$1$ and
lag-$12$ correlations across subcarriers ($24$); the mean absolute
adjacent-subcarrier difference and the across-subcarrier SD of the
symbol-averaged $\log|y|^2$ add two. For
mobility, supervised controls were trained on the CPU in float32, with
learning rates selected separately for each seed and the same epoch caps.

\paragraph{Budgets and uncertainty}
A nominal budget of $B$ labels uses $0.8B$ for fitting and $0.2B$ for
validation, with indices shared by all methods. The full budget is
$B=10{,}000$, and the matched low-label budgets are $100$, $500$, and
$1{,}000$. Every learned model, including each pre-trained variant, is trained
with seeds $0$, $2$, and $3$, and we report the mean $\pm$ sample standard
deviation. For the predeclared primary contrasts we give crossed $95\%$
bootstrap intervals that resample the three training seeds and the
generation batches; they are neither equivalence intervals nor population
guarantees. Ablation intervals are paired batch-bootstrap intervals
given seed~$0$.}

\subsection{Main Results}
\label{sec:mainresults}

\begin{figure}[!t]
\centering
\includegraphics[width=\columnwidth]{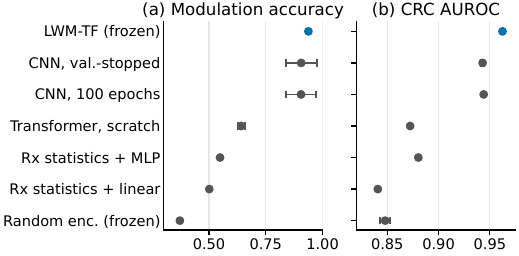}
\caption{Frozen LWM-TF with a linear probe against the controls on test set~1 at the
full budget ($8{,}000$ fitting, $2{,}000$ validation labels). Mean $\pm$ SD over
three seeds; the Rx-statistics linear probe is a single fit.}
\label{fig:controls}
\end{figure}

\begin{table}[!t]
\caption{Primary contrasts at the full budget.}
\label{tab:primary}
\centering
\footnotesize
\setlength{\tabcolsep}{3.5pt}
\begin{tabular}{@{}llccc@{}}
\toprule
Task & Test set & P0 (frozen) & Stat.\ MLP & CNN (ext.)\\
\midrule
CRC  & 1 & $.9623\pm.0012$ & $.8803\pm.0009$ & $.9429\pm.0016$\\
     &       &                 & $[.0743,.0893]$ & $[.0150,.0235]$\\
CRC  & 2 & $.9642\pm.0016$ & $.8888\pm.0012$ & $.9446\pm.0024$\\
     &       &                 & $[.0646,.0862]$ & $[.0135,.0256]$\\
\addlinespace[2pt]
Mod. & 1 & $.9377\pm.0024$ & $.5486\pm.0002$ & $.9058\pm.0678$\\
     &       &                 & $[.3833,.3949]$ & $[-.0081,.1073]$\\
Mod. & 2 & $.9387\pm.0013$ & $.5503\pm.0013$ & $.9060\pm.0676$\\
     &       &                 & $[.3815,.3951]$ & $[-.0078,.1083]$\\
\bottomrule
\end{tabular}
\end{table}

\begin{figure}[!t]
\centering
\includegraphics[width=\columnwidth]{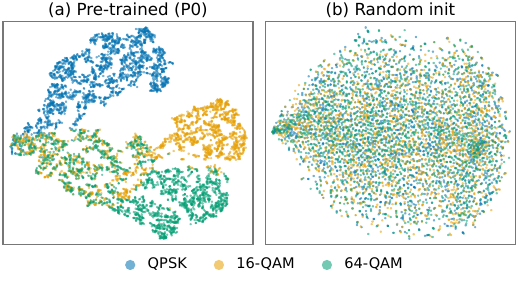}
\caption{t-SNE of frozen embeddings of $5{,}400$ test-set-2 slots, colored by
modulation: (a) P0, (b) random-init encoder.}
\label{fig:tsne}
\end{figure}

{Fig.~\ref{fig:controls} compares the main model (P0) with
every control on test set~1, and Table~\ref{tab:primary} gives the
predeclared contrasts on both test sets at $10{,}000$ labels; its bracketed rows
give the crossed $95\%$ interval of P0 minus each column's control.

\emph{SNR.} SNR is not plotted in Fig.~\ref{fig:controls}. The parameter-free
estimator reaches $R^2=0.9992$ and the receiver-statistic probes
$0.9992$--$0.9993$, above the frozen encoder's $0.9990$ and the CNN's $0.9903$. SNR is solved in closed form on this input, and we do not count it as
evidence for the learned representation.

\emph{CRC decodability.} The frozen encoder exceeds the receiver-statistic MLP
by $0.075$--$0.082$ AUROC and the evaluated CNN by $0.019$--$0.020$ on the
two test sets; all four crossed intervals exclude zero. The CNN is trained
without data augmentation, so we treat its margin as specific to this recipe.

\emph{Modulation classification.} The frozen encoder exceeds the
receiver-statistic MLP by $0.39$ on both test sets, the task on which the seven
statistics fail most clearly. It is $0.032$--$0.033$ above the evaluated CNN, but
the crossed intervals include zero: the CNN's three seeds reach $0.828$,
$0.944$, and $0.946$ on test set~2, against $0.937$--$0.940$ for P0. The low
CNN seed stopped by validation patience at epoch $152$ of $400$. We retain
all three runs; at the full budget the two are on par here.

The randomly-initialized encoder, read out by the same probe, stays near
chance on modulation ($0.3717$) but reaches $0.8479$ on CRC, consistent with
its SNR $R^2$ of $0.6152$: CRC outcomes follow link quality, which random
features partly retain. The scratch transformer, trained end to end at the same budget,
reaches $0.6416$ and $0.8723$. These two comparisons are descriptive and were
not among the predeclared contrasts.

\emph{Secondary diagnostics.} On test set~1, with the random encoder
and the receiver-statistic linear probe in parentheses, the frozen probe reads
the DMRS position almost perfectly (accuracy $0.998$; $0.787$, $0.506$), as
expected for a pilot pattern visible in the input, so this task serves as a
positive control. LoS AUROC is $0.820$ ($0.743$, $0.798$), a small margin over
the receiver statistics, and scenario accuracy is $0.575$ ($0.422$, $0.449$)
against a chance level of $0.33$. Speed reaches $R^2=0.529$ ($0.047$, $0.001$)
and $0.46$--$0.59$ within each scenario, so Doppler information from a single
slot is retained; Section~\ref{sec:mobility} evaluates it as a task. The
pooled distance $R^2$ of $0.293$ ($0.050$, $0.058$) does not indicate ranging:
applied within each scenario, the same probe has negative $R^2$, and a probe
refitted within each scenario reaches only $R^2\approx-0.06$, so the pooled
score reflects the scenarios' distance ranges, as expected when neither the
received power (normalized away) nor the absolute delay (TR~38.901 delays start
at the first cluster) reaches the input.}

The frozen encoder ($8.46$\,M parameters, $11.89$\,GMAC per slot) runs
efficiently on batched slots: on an NVIDIA A100 in bfloat16 it processes
$1{,}431$ slots/s at batch $32$ ($0.70$\,ms per slot), whereas a single slot
takes $5.18$\,ms. It therefore suits batched and non-real-time analysis, such
as offline monitoring of recorded grids, in which one encoder pass can serve
several probes; per-slot processing under a tight latency budget favors a
smaller task-specific model.

{Fig.~\ref{fig:tsne} visualizes the frozen embeddings of the
seed-0 encoders with t-SNE~\cite{vandermaaten2008tsne} (features standardized and reduced to 50 dimensions
by principal component analysis; perplexity 40). The pre-trained encoder separates the three
modulation orders into distinct regions, with 16- and 64-QAM meeting along one
boundary band, whereas the random-init encoder shows no modulation structure.
The contrast holds before projection: in the 50-dimensional space, the 20
nearest neighbors of a slot share its modulation order $85\%$ of the time for
P0 and $33\%$, the chance level, for the random-init encoder. On test set~2
the P0 probe recalls QPSK, 16-QAM, and 64-QAM at $99.0\%$, $91.0\%$, and
$91.6\%$ (three-seed means), and $91\%$ of its errors confuse
the two QAM orders, the same residual confusion as the evaluated CNN. The
projection is qualitative; quantitative claims rest on the probe results
above.}

\subsection{Channel Time-Variation Detection}
\label{sec:mobility}

\begin{table}[!t]
\caption{Mobility detection (AUROC; mean $\pm$ SD over seeds).}
\label{tab:mobility}
\centering
\footnotesize
\setlength{\tabcolsep}{2pt}
\begin{tabular}{@{}rccccc@{}}
\toprule
Labels & P0 (frozen) & CNN & Stat.\ MLP & Doppler & Random\\
\midrule
$100$     & $.786\pm.017$ & $.496\pm.003$ & $.489\pm.004$ & $.481\pm.002$ & $.503\pm.003$\\
$500$     & $.856\pm.011$ & $.496\pm.005$ & $.504\pm.002$ & $.556\pm.058$ & $.526\pm.009$\\
$1{,}000$ & $.890\pm.004$ & $.497\pm.006$ & $.512\pm.007$ & $.498\pm.005$ & $.534\pm.009$\\
$10{,}000$& $.916\pm.003$ & $.498\pm.003$ & $.758\pm.002$ & $.817\pm.005$ & $.580\pm.011$\\
\midrule
\multicolumn{6}{@{}l}{P0 (frozen), $10{,}000$ labels:}\\
\multicolumn{4}{@{}l}{\quad mean within the $12$ settings} & \multicolumn{2}{c}{$.916\pm.003$}\\
\multicolumn{4}{@{}l}{\quad without stationary slots} & \multicolumn{2}{c}{$.894\pm.002$}\\
\multicolumn{4}{@{}l}{Predictor using only the setting} & \multicolumn{2}{c}{$.499$}\\
\bottomrule
\end{tabular}
\end{table}

{Table~\ref{tab:mobility} reports the channel time-variation task on test
set~2 (mean $\pm$ sample SD over three seeds; chance $0.5$; test set~1 within
$0.012$), with the Doppler features of Section~\ref{sec:protocol} as the
task-specific control.
Up to $1{,}000$ labels every supervised and engineered control stays near
chance, and the CNN stays at chance even with $10{,}000$ labels. The frozen
probe reaches $0.79$ with $100$ labels, $0.89$ with $1{,}000$, and $0.92$ with
$10{,}000$, where it also exceeds the task-specific Doppler features by $0.099$
$[0.088,0.111]$. All six predeclared intervals against the CNN and the
statistics MLP at $100$--$1{,}000$ labels exclude zero on both test sets, and
the frozen probe wins all five label re-draws at every budget. The lower block
of Table~\ref{tab:mobility} rules out a shortcut through the generation
setting: the score holds within each scenario--DMRS setting and stays well
above chance without the stationary slots, whereas a predictor that knows
only the setting, through its mobile rate in the pre-training corpus, is at
chance. Mobility is the task on which pre-training adds what neither
supervision on the grid nor hand-built channel statistics fully recover: the
channel's variation across the OFDM symbols of a slot, which the grid CNN does
not discover from labels.}

\subsection{Label Efficiency}
\label{sec:labeleff}

\begin{figure}[!t]
\centering
\includegraphics[width=\columnwidth]{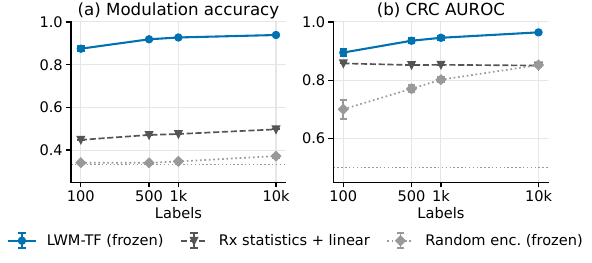}
\caption{Label efficiency on test set~2: the same linear probe on three
feature sets, fitted on $80\%$ of each nominal budget. Mean $\pm$ SD over
three seeds (receiver statistics are deterministic); dotted: chance; test
set~1 within $0.01$.}
\label{fig:labeleff}
\end{figure}
\begin{figure*}[!t]
\centering
\includegraphics[width=\textwidth]{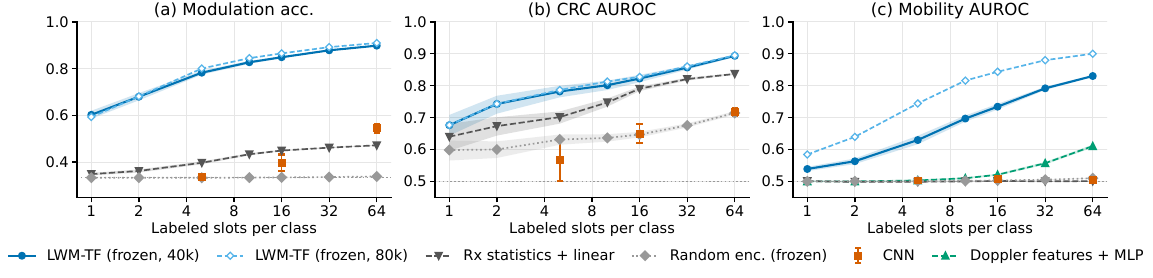}
\caption{$K$ labeled slots per class on test set~2: mean over $100$ nested
support-set draws (CNN: $5$ draws at $K=5$, $16$, $64$); bands and bars are
$95\%$ bootstrap intervals; dotted: chance. Dashed: the encoder continued to
$80{,}000$ steps (Fig.~\ref{fig:horizon}), scored after the other results.}
\label{fig:kshot}
\end{figure*}
\begin{figure}[!t]
\centering
\includegraphics[width=\columnwidth]{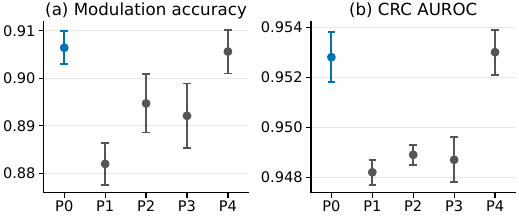}
\caption{Objective and masking ablations on test set~1, all models at $5{,}000$ steps.
P0: log-power reconstruction and scale target (main objective); P1: no scale
target; P2: no reconstruction; P3: complex-I/Q reconstruction; P4:
DMRS-stratified masking. Mean $\pm$ SD over three seeds.}
\label{fig:ablation}
\end{figure}

{Label counts here refer to downstream adaptation; pre-training uses none.
Fig.~\ref{fig:labeleff} shows how the frozen probe behaves as labels become
scarce. With $100$ labels it reaches $0.87$ modulation accuracy and $0.89$ CRC
AUROC, within $0.07$ of its scores with $10{,}000$ labels, and $0.79$ mobility
AUROC (Table~\ref{tab:mobility}). Read out by the same probe, the seven
receiver statistics reach $0.45$ and $0.86$ with $100$ labels and gain little
with more, and the randomly-initialized encoder stays near chance on
modulation. The pre-trained features lead at every budget and on every seed,
so the gain comes from pre-training rather than from the probe. These two
curves were added after the primary results, with the probe protocol
unchanged.

\emph{Few labels per class.} Fig.~\ref{fig:kshot} counts labels
per class instead: $K$ labeled slots per class from the development pool, with
features standardized by the unlabeled pool so that a probe uses exactly $K$
labels per class and no validation labels; supervised controls train on the
support set with AdamW (rate $10^{-3}$, weight decay $0.01$, full batch,
clipping at $1$) for a fixed $200$ epochs without validation. Each of $100$ draws orders every class once and takes its
first $K$ slots, so the support sets of a draw are nested across $K$. The CNN
uses the first five of those draws at $K=5$, $16$, and $64$. With $K=16$ the
frozen probe reaches $0.85$ modulation accuracy, $0.82$ CRC AUROC,
and $0.73$ mobility AUROC; the seed-0 continuation to $80{,}000$ steps
(dashed) reaches $0.87$, $0.83$, and $0.84$. The random-init encoder stays at chance on
modulation and mobility; the receiver statistics reach $0.45$, $0.79$, and
$0.50$, and the Doppler-feature MLP $0.52$ on mobility. On the five support
draws shared with the CNN, the frozen probe averages $0.86$, $0.86$, and
$0.73$, against CNN scores of $0.40$, $0.65$, and $0.51$; it leads on every
paired draw in all three tasks.
The CRC probe depends most on the support set (SD over draws $0.17$ at $K=1$,
$0.06$ at $K=16$), yet its mean exceeds the receiver statistics at every $K$:
by $0.03$ [$0.02$, $0.04$] at $K=16$ and by $0.06$ [$0.05$, $0.06$] at
$K=64$, where it wins all $100$ draws. This analysis was designed after the
primary results; its protocol was fixed before scoring and amended once, to
nested supports and $100$ draws, after independent draws per $K$ zig-zagged
within their sampling error.}

\begin{remark}[When does the frozen encoder help?]
The frozen probes retain modulation, CRC, and mobility information with
few labels. Against the evaluated CNN, the frozen probe leads on
every paired draw with $16$ labeled slots per class, whereas at the full budget
the differences are small and specific to the CNN recipe. One encoder pass can serve several tasks and runs
efficiently on batched slots, whereas a single latency-critical task favors
the cheaper CNN, and SNR favors the closed-form estimator.
\end{remark}
\begin{table}[!t]
\caption{Held-out scenario and DMRS configuration.}
\label{tab:generalization}
\centering
\footnotesize
\setlength{\tabcolsep}{1.8pt}
\begin{tabular}{@{}lcccc@{}}
\toprule
 & \multicolumn{4}{c}{Held out}\\
\cmidrule(l){2-5}
Model & None & RMa & DMRS~3 & Both\\
\midrule
\multicolumn{5}{@{}l}{\emph{Modulation (accuracy)}}\\
D0 (frozen) & $.9047\pm.0005$ & $.9054\pm.0039$ & $.8229\pm.0572$ & $.8251\pm.0502$\\
CNN         & $.8746\pm.0630$ & $.8789\pm.0581$ & $.7857\pm.1335$ & $.7924\pm.1301$\\
Stat.\ MLP  & $.5758\pm.0019$ & $.5127\pm.0013$ & $.5193\pm.0011$ & $.5125\pm.0013$\\
\addlinespace[2pt]
\multicolumn{5}{@{}l}{\emph{CRC decodability (AUROC)}}\\
D0 (frozen) & $.9454\pm.0008$ & $.9403\pm.0024$ & $.9657\pm.0007$ & $.9788\pm.0014$\\
CNN         & $.9382\pm.0020$ & $.9419\pm.0015$ & $.9628\pm.0032$ & $.9768\pm.0056$\\
Stat.\ MLP  & $.8642\pm.0002$ & $.8614\pm.0019$ & $.9167\pm.0021$ & $.9591\pm.0003$\\
\bottomrule
\end{tabular}
\end{table}
\begin{table}[!t]
\caption{Anomaly detection without impairment labels (AUROC).}
\label{tab:anomaly}
\centering
\footnotesize
\setlength{\tabcolsep}{2.5pt}
\begin{tabular}{@{}lccccccc@{}}
\toprule
& \multicolumn{2}{c}{NBI} & \multicolumn{2}{c}{Burst} & \multicolumn{2}{c}{Sat.+quant.} & \\
\cmidrule(lr){2-3}\cmidrule(lr){4-5}\cmidrule(lr){6-7}
Model & Mild & Severe & Mild & Severe & Mild & Severe & Macro\\
\midrule
P0 (frozen)    & $.978$ & $.999$ & $.606$ & $1.000$ & $.506$ & $.594$ & $.780$\\
Random encoder & $.653$ & $.991$ & $.501$ & $.818$ & $.499$ & $.499$ & $.660$\\
Rx statistics  & $.554$ & $.984$ & $.665$ & $.997$ & $.474$ & $.400$ & $.679$\\
\bottomrule
\end{tabular}
\end{table}
\begin{figure}[!t]
\centering
\includegraphics[width=\columnwidth]{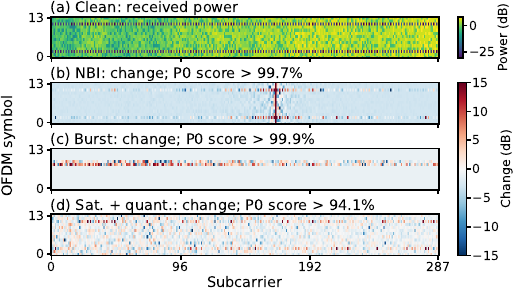}
\caption{Severe impairments on one validation link (antenna~0; median SNR,
$22.3$\,dB, of the $140$ in $[20,25)$\,dB): (a) clean power; (b)--(d) change
(RMS normalization lowers unaffected REs). Titles: P0 clean-only score
percentile.}
\label{fig:anomalyex}
\end{figure}

\subsection{Objective and Masking Ablations}
\label{sec:ablation}

Fig.~\ref{fig:ablation} separates the components of the
objective at a matched $5{,}000$-step budget. \emph{Reconstruction and the scale target are complementary}:
removing the scale target (P1) costs $0.024$ in modulation accuracy and
$0.005$ in CRC AUROC, and removing reconstruction (P2) costs $0.012$ and
$0.004$; all four seed-0 paired intervals exclude zero, and neither component
alone reaches the combination. The scale target alone is already a strong
pretext, so we do not attribute all learned structure to reconstruction.
\emph{Log-power outperforms complex reconstruction}: the complex-I/Q model (P3)
regresses the $24$ I/Q values of each masked patch of the RMS-normalized grid
by mean squared error, without per-patch normalization and with P0's scale
terms, and trails P0 by $0.014$ and $0.004$, with both paired intervals excluding zero.
It nevertheless remains far above the random encoder ($0.8921$ against
$0.3717$ modulation accuracy), so the partial degeneracy of
\eqref{eq:degenerate} makes the complex objective less efficient, not
unusable. \emph{DMRS-stratified masking performs on par with uniform masking}: P4 and P0 differ by at
most $0.001$ on both tasks, and both paired intervals include zero.

{\emph{Relative-phase supervision improves spatial readout but costs
modulation and speed performance.} An additional model, P5, adds
\eqref{eq:lphi} to P0 at $\beta_{\varphi}=1$. P0 and P5 use the same dataset
and schedule, with three seed-matched runs. At $40{,}000$ steps on
test set~1, P5 raises LoS AUROC from $0.820$ to $0.872$
($+0.101$ in RMa) and scenario accuracy from $0.575$ to $0.618$, but lowers
modulation accuracy from $0.938$ to $0.911$, CRC AUROC from $0.962$ to
$0.953$, and speed $R^2$ from $0.529$ to $0.395$. At $5{,}000$,
$10{,}000$, and $20{,}000$ steps, P5 also trails P0 on speed $R^2$ by
$0.050$, $0.087$, and $0.107$, respectively; both improve with training.
The same directions hold at $5{,}000$ steps for LoS
($0.812\rightarrow0.856$) and modulation ($0.906\rightarrow0.875$),
the horizon used for the P0--P4 ablations in Fig.~\ref{fig:ablation}.
P5 was designed after test set~2 was generated and has not been scored on
that set, so we report these comparisons descriptively.

A separate $5{,}000$-step diagnostic tests whether the gain is spatial.
Ridge probes fitted on $64$ generation batches of test set~1 and scored on
the other $320$ explain $59\%$ of the variance of a label-free
dominant-eigenvalue index of the noise-corrected port covariance with P5
features, against $47\%$ with P0 ($32\%$ against $19\%$ after removing SNR,
LoS, and scenario). A frequency-selectivity control moves only from
$0.62$ to $0.65$. The phase target therefore adds inter-port structure;
its modulation cost is consistent with the target's removal of payload
amplitude. Because $\beta_{\varphi}$ was not tuned, the costs may also reflect
the loss weight or competition between targets. The main model keeps
$\beta_{\varphi}=0$.}

\subsection{Held-Out Scenario and DMRS Configuration}
\label{sec:generalization}

Table~\ref{tab:generalization} divides test set~1 by the condition held out
from the source-only models: none, the RMa scenario, DMRS position~3, or both
(mean $\pm$ SD over three seeds).
The encoder is pre-trained for $5{,}000$ steps, and its probe, the CNN (up to
$100$ epochs), and the receiver-statistic MLP are fitted and validated on the source
conditions only. RMa leaves modulation accuracy near $0.905$, whereas the
unseen DMRS configuration lowers it to $0.823$--$0.825$, with substantial
seed variability. CRC AUROC changes little under RMa and rises in the
held-DMRS columns. Each column has a different test distribution, so these
differences reflect the columns as well as the shift. The same probes transfer
SNR ($R^2$ of $0.997$--$0.998$ in every column) but not LoS in RMa (AUROC
$0.61$--$0.63$, against $0.66$--$0.67$ for P0 at $5{,}000$ steps) or speed
when both settings are held out ($R^2=0.01$, against $0.21$).

Continued to $40{,}000$ steps, the three source-only encoders give modulation
accuracy of $0.941/0.939/0.930/0.921$ and CRC AUROC of
$0.961/0.958/0.972/0.985$ in the four columns. Against the same-seed P0 at the
same step, scored on the same slots, the modulation gap in the held-DMRS
columns shrinks from $0.075$/$0.077$ at $5{,}000$ steps to $0.005$/$0.007$ (at
most $0.013$ per seed); CRC AUROC is within $0.005$. Speed with both held out
reaches $R^2=0.53$ (P0: $0.47$), but LoS in RMa stays $0.05$ AUROC below P0.
These results concern held-out conditions within TR~38.901.

\subsection{Waveform Anomaly Detection}
\label{sec:anomaly}

Table~\ref{tab:anomaly} reports detection on the $2{,}048$ test parent links by
severity (mean over three seeds, SD $\le 0.03$). No impairment is seen: each
feature space models $2{,}000$ clean fit-split grids as a Gaussian and scores a
test grid by its Mahalanobis distance~\cite{lee2018mahalanobis} (clean-statistics
standardization; covariance $\mathbf{S}$ shrunk to
$0.9\mathbf{S}+0.1(\operatorname{tr}\mathbf{S}/d)\mathbf{I}$ for $d$ features). The frozen
encoder is best (macro $0.780\pm0.005$), and its margin lies in mild NBI
(Fig.~\ref{fig:anomalyex}): $0.978$ against $0.554$ for the statistics, which
are slightly better on mild bursts ($0.665$ against $0.606$); severe NBI and
bursts are near $1$ for both. Saturation, whose distortion resembles in-band white noise,
stays near chance when mild; when severe, P0 reaches $0.594$ ($0.845$ above
$22$\,dB apparent SNR) and the statistics $0.400$.

This clean-only protocol was added after a held-family protocol, which trains
each detector on the other two families, and was fixed before test scoring. In
the held-family protocol P0 shows no advantage: macro AUROC is $0.613$ against
$0.637$ for the random encoder, $0.567$ for the CNN, and $0.519$ for the
statistics MLP, and P0's seeds disagree on NBI and burst (SD $0.13$ and
$0.28$). The representation thus flags off-manifold grids, but a head trained
on some impairment families does not transfer to a held-out one. The
impairments are parametric models; measured interference and converter effects
are a natural extension.

\section{Conclusion}
\label{sec:conclusion}
{We presented LWM-TF, a $9.58$\,M-parameter masked autoencoder
pre-trained on the 5G NR received resource grid after the FFT and before
channel estimation. It learns by reconstructing the log-power of masked PRB
tokens together with a per-patch scale target; controlled ablations show that
both terms contribute and that log-power is a more effective target than the
complex grid. A single frozen encoder with linear probes reaches $0.939$
modulation accuracy, $0.964$ CRC AUROC, and $0.92$ mobility AUROC with
$10{,}000$ labels on a separately generated test set, well above engineered
receiver statistics. With $16$ labeled slots per class it leads a supervised
CNN on every paired draw, it captures channel time variation that the CNN does
not learn from labels, and it flags unseen waveform impairments without
supervision. Validating these gains on measured data and reducing the encoder
cost for real-time use are the next steps.}

\appendices
\section{Waveform-Domain Anomaly Injection}
\label{app:anomaly}
This appendix details the waveform generation and detector protocols of
Section~\ref{sec:anomaly}.
Each parent link of the anomaly experiment is generated as a
time-domain waveform rather than taken from the dataset. A PUSCH slot, with the
modulation drawn uniformly and the MCS drawn uniformly within it, is mapped to
the $288$ occupied bins of a $512$-point FFT and OFDM-modulated with a
$48$-sample cyclic prefix at $30$\,kHz subcarrier spacing. It passes through a
time-varying TR~38.901 channel (the twelve scenario--DMRS combinations are
cycled; every fourth batch is stationary, otherwise the speed is drawn from
$\mathcal{U}(0,120)$\,km/h), and complex additive white Gaussian noise is added at a target SNR drawn
from $\mathcal{U}(-5,30)$\,dB relative to the clean grid power. The receiver
normalizes the four-antenna waveform to unit RMS (AGC), injects at most one
impairment, removes the CP, applies the FFT, keeps the occupied bins, and
RMS-normalizes the grid as in Section~\ref{sec:crclabel}, so total power
carries no label information.

With $z_a[n]$ the AGC-normalized waveform of antenna $a$, the three families
and their two severity levels are as follows.
\emph{NBI}: $z_a[n]+10^{I/20}e^{\jmath(2\pi\nu n+\phi_a)}$, a tone at a
normalized frequency $\nu$ drawn uniformly within the occupied band, with an
independent phase $\phi_a$ per antenna and interference level
$I\in\{-10,0\}$\,dB relative to the normalized received power.
\emph{Burst}: complex Gaussian noise of power $10^{I/10}$ added over a
contiguous window covering $2\%$ or $5\%$ of the slot's samples, with
$I=0$ or $10$\,dB, respectively.
\emph{Saturation and quantization}: the I and Q components are clipped to
$\pm A/\sqrt{2}$ and uniformly quantized over that range to $b$ bits, with
$(A,b)=(2.5,6)$ or $(1.5,4)$; we call this an analog-to-digital-converter-like impairment, not a model
of a specific converter. Each parent is stored with its clean grid and six
impaired variants, all in one split; the splits hold $3{,}504$, $880$, and
$2{,}048$ parent links for fitting, validation, and test.

{In the held-family protocol, each detector is fitted on $8{,}000$ samples and
validated on $2{,}000$ ($20\%$ clean, $80\%$ from the two source families): a
linear head on the frozen encoders, the MLP of Section~\ref{sec:protocol} on
the receiver statistics, or the CNN end to end, with the learning rate chosen
on source-family validation loss; no test sample is used for fitting,
selection, or score orientation.}

\section{Receiver Statistics}
\label{app:aux}
This appendix defines the seven receiver statistics used as a control in
Section~\ref{sec:downstream}.
{The seven statistics use only the received grid and the fixed
receiver configuration. Let $p_a[t,f]=|Y_a[t,f]|^2$ after RMS normalization.
Symbol $2$ carries DMRS in every setting, and with one DMRS port its odd
subcarriers are empty, so their mean power $P_{\mathrm{n}}$ measures the noise
directly; its even subcarriers give the pilot power $P_{\mathrm{p}}$. The nine
symbols $\{0,1,3,4,6,9,10,12,13\}$ carry no DMRS in any setting and give the
per-antenna data power $P_{\mathrm{d},a}$, with mean $P_{\mathrm{d}}$ over
antennas; $P$ is the mean of $p_a[t,f]$ over the whole grid. The statistics
are the mean and the variance of $\log p_a[t,f]$ over all REs and antennas;
$\log(P_{\mathrm{n}}/P)$, $\log(P_{\mathrm{d}}/P)$, and
$\log(P_{\mathrm{p}}/P)$; the SNR estimate
$\hat{\rho}=10\log_{10}\big((P_{\mathrm{d}}-P_{\mathrm{n}})/P_{\mathrm{n}}\big)$;
and the variance of $\log P_{\mathrm{d},a}$ across antennas. The
parameter-free SNR estimator of Section~\ref{sec:mainresults} is $\hat{\rho}$ itself,
with no fitted parameter.}



\begin{thebibliography}{99}


\bibitem[1]{chen2024bigai}
Z.~Chen, Z.~Zhang, and Z.~Yang, ``Big AI models for 6G wireless networks:
Opportunities, challenges, and research directions,'' \emph{IEEE Wireless Commun.}, vol.~31, no.~5, pp. 164--172, 2024.

\bibitem[2]{oshea2017intro}
T.~O'Shea and J.~Hoydis, ``An introduction to deep learning for the physical
layer,'' \emph{IEEE Trans. Cogn. Commun. Netw.},
vol.~3, no.~4, pp. 563--575, 2017.

\bibitem[3]{aitaoudia2022e2e}
F.~Ait~Aoudia and J.~Hoydis, ``End-to-end learning for OFDM: From neural
receivers to pilotless communication,'' \emph{IEEE Trans. Wireless Commun.},
vol.~21, no.~2, pp. 1049--1063, Feb. 2022.

\bibitem[4]{devlin2019bert}
J.~Devlin, M.-W. Chang, K.~Lee, and K.~Toutanova, ``BERT: Pre-training of deep
bidirectional transformers for language understanding,'' in \emph{Proc. NAACL-HLT}, 2019, pp. 4171--4186.

\bibitem[5]{brown2020gpt3}
T.~B. Brown \emph{et al.}, ``Language models are few-shot learners,''
\emph{Adv. Neural Inf. Process. Syst.}, vol.~33, pp.
1877--1901, 2020.

\bibitem[6]{he2022mae}
K.~He, X.~Chen, S.~Xie, Y.~Li, P.~Doll\'{a}r, and R.~Girshick, ``Masked
autoencoders are scalable vision learners,'' in \emph{Proc. IEEE/CVF CVPR}, 2022, pp.
16\,000--16\,009.

\bibitem[7]{baevski2020wav2vec2}
A.~Baevski, Y.~Zhou, A.~Mohamed, and M.~Auli, ``wav2vec 2.0: A framework for
self-supervised learning of speech representations,'' \emph{Adv. Neural Inf. Process. Syst.}, vol.~33, pp. 12\,449--12\,460, 2020.

\bibitem[8]{alikhani2025lwm}
S.~Alikhani, G.~Charan, and A.~Alkhateeb, ``Large wireless model (LWM): A
foundation model for wireless channels,'' \emph{arXiv preprint
arXiv:2411.08872}, 2025.

\bibitem[9]{liu2025wifo}
B.~Liu, S.~Gao, X.~Liu, X.~Cheng, and L.~Yang, ``WiFo: Wireless foundation
model for channel prediction,'' \emph{Sci. China Inf. Sci.},
vol.~68, no.~6, p. 162302, 2025.

\bibitem[10]{guler2026contrawimae}
B.~Guler, G.~Geraci, and H.~Jafarkhani, ``A multi-task foundation model for
wireless channel representation using contrastive and masked autoencoder
learning,'' \emph{IEEE J. Sel. Areas Commun.}, vol.~44,
pp. 4489--4504, 2026.

\bibitem[11]{guler2026pilotwimae}
B.~Guler, G.~Geraci, and H.~Jafarkhani, ``PilotWiMAE: Pilot-native representation
learning for wireless channels,'' \emph{arXiv preprint arXiv:2605.22856}, 2026.

\bibitem[12]{ott2024radiofm}
J.~Ott, J.~Pirkl, M.~Stahlke, T.~Feigl, and C.~Mutschler, ``Radio foundation
models: Pre-training transformers for 5G-based indoor localization,'' in
\emph{Proc. Int. Conf. Indoor Position. Indoor Navig. (IPIN)}, Kowloon, Hong Kong, 2024, pp. 1--6.

\bibitem[13]{liu2026wifocf}
X.~Liu, S.~Gao, B.~Liu, X.~Cheng, and L.~Yang, ``WiFo-CF: Wireless foundation
model for CSI feedback,'' \emph{IEEE Trans. Wireless Commun.}, vol.~25, pp. 15\,039--15\,053, 2026.

\bibitem[14]{liu2024llm4cp}
B.~Liu, X.~Liu, S.~Gao, X.~Cheng, and L.~Yang, ``LLM4CP: Adapting large
language models for channel prediction,'' \emph{J. Commun. Inf. Netw.}, vol.~9, no.~2, pp. 113--125, June 2024.

\bibitem[15]{aboulfotouh2024radio}
A.~Aboulfotouh, A.~Eshaghbeigi, and H.~Abou-Zeid, ``Building 6G radio
foundation models with transformer architectures,'' \emph{arXiv preprint
arXiv:2411.09996}, 2024.

\bibitem[16]{kim2026lwmspectro}
N.~Kim, S.~Alikhani, and A.~Alkhateeb, ``LWM-Spectro: A foundation model for
wireless baseband signal spectrograms,'' \emph{arXiv preprint
arXiv:2601.08780}, 2026.

\bibitem[17]{mashaal2026iqfm}
O.~Mashaal and H.~Abou-Zeid, ``IQFM---a wireless foundation model for I/Q
streams in AI-native 6G,'' \emph{IEEE Open J. Commun. Soc.}, vol.~7, pp. 1426--1441, 2026.

\bibitem[18]{aboulfotouh2025wavesfm}
A.~Aboulfotouh, E.~Mohammed, and H.~Abou-Zeid, ``6G WavesFM: A foundation model
for sensing, communication, and localization,'' \emph{IEEE Open J.
Commun. Soc.}, vol.~6, pp.~6792--6807, 2025,
doi:~10.1109/\allowbreak OJCOMS.2025.3600616.

\bibitem[19]{mohamed2026latentwave}
A.~Mohamed, A.~Aboulfotouh, and H.~Abou-Zeid, ``LatentWave: JEPA pretraining
for wireless foundation models,'' \emph{arXiv preprint arXiv:2606.06373}, 2026.

\bibitem[20]{honkala2021deeprx}
M.~Honkala, D.~Korpi, and J.~M.~J. Huttunen, ``DeepRx: Fully convolutional deep
learning receiver,'' \emph{IEEE Trans. Wireless Commun.},
vol.~20, no.~6, pp. 3925--3940, 2021.

\bibitem[21]{ye2018power}
H.~Ye, G.~Y. Li, and B.-H. Juang, ``Power of deep learning for channel
estimation and signal detection in OFDM systems,'' \emph{IEEE Wireless Commun. Lett.}, vol.~7, no.~1, pp. 114--117, 2018.

\bibitem[22]{oshea2018ota}
T.~J. O'Shea, T.~Roy, and T.~C. Clancy, ``Over-the-air deep learning based
radio signal classification,'' \emph{IEEE J. Sel. Topics Signal Process.}, vol.~12, no.~1, pp. 168--179, 2018.

\bibitem[23]{wen2018csinet}
C.-K. Wen, W.-T. Shih, and S.~Jin, ``Deep learning for massive MIMO CSI
feedback,'' \emph{IEEE Wireless Commun. Lett.}, vol.~7, no.~5, pp.
748--751, 2018.

\bibitem[24]{alkhateeb2018coordinated}
A.~Alkhateeb, S.~Alex, P.~Varkey, Y.~Li, Q.~Qu, and D.~Tujkovic, ``Deep
learning coordinated beamforming for highly-mobile millimeter wave systems,''
\emph{IEEE Access}, vol.~6, pp. 37\,328--37\,348, 2018.

\bibitem[25]{chen2020simclr}
T.~Chen, S.~Kornblith, M.~Norouzi, and G.~Hinton, ``A simple framework for
contrastive learning of visual representations,'' in \emph{Proc. Int. Conf. Mach. Learn. (ICML)}, 2020, pp.
1597--1607.

\bibitem[26]{assran2023ijepa}
M.~Assran, Q.~Duval, I.~Misra, P.~Bojanowski, P.~Vincent, M.~Rabbat, Y.~LeCun,
and N.~Ballas, ``Self-supervised learning from images with a joint-embedding
predictive architecture,'' in \emph{Proc. IEEE/CVF CVPR}, 2023, pp. 15\,619--15\,629.

\bibitem[27]{3gpp38901}
3GPP, ``Study on channel model for frequencies from 0.5 to 100 GHz,'' 3GPP TR 38.901, V17.0.0,
2022.

\bibitem[28]{hoydis2022sionna}
J.~Hoydis, S.~Cammerer, F.~Ait~Aoudia, A.~Vem, N.~Binder, G.~Marcus, and
A.~Keller, ``Sionna: An open-source library for next-generation physical layer
research,'' \emph{arXiv preprint arXiv:2203.11854}, 2022.

\bibitem[29]{3gpp38211}
3GPP, ``NR; physical channels and modulation,'' 3GPP TS 38.211, V17.4.0, 2023.

\bibitem[30]{3gpp38212}
------, ``NR; multiplexing and channel coding,'' 3GPP TS 38.212, V17.4.0, 2023.

\bibitem[31]{3gpp38214}
------, ``NR; physical layer procedures for data,'' 3GPP TS 38.214, V17.4.0, 2023.

\bibitem[32]{alkhateeb2019deepmimo}
A.~Alkhateeb, ``DeepMIMO: A generic deep learning dataset for millimeter wave
and massive MIMO applications,'' in \emph{Proc. Inf. Theory Appl. Workshop
(ITA)}, San Diego, CA, USA, Feb. 2019, pp. 1--8.

\bibitem[33]{bertasius2021timesformer}
G.~Bertasius, H.~Wang, and L.~Torresani, ``Is space-time attention all you
need for video understanding?'' in \emph{Proc. Int. Conf. Mach. Learn.
(ICML)}, 2021, pp. 813--824.

\bibitem[34]{strodthoff2019harq}
N.~Strodthoff, B.~G\"{o}ktepe, T.~Schierl, C.~Hellge, and W.~Samek, ``Enhanced
machine learning techniques for early HARQ feedback prediction in 5G,''
\emph{IEEE J. Sel. Areas Commun.}, vol.~37, no.~11, pp.
2573--2587, Nov. 2019.

\bibitem[35]{vandermaaten2008tsne}
L.~van der Maaten and G.~Hinton, ``Visualizing data using t-SNE,''
\emph{J. Mach. Learn. Res.}, vol.~9, no.~86, pp. 2579--2605, 2008.

\bibitem[36]{lee2018mahalanobis}
K.~Lee, K.~Lee, H.~Lee, and J.~Shin, ``A simple unified framework for
detecting out-of-distribution samples and adversarial attacks,''
\emph{Adv. Neural Inf. Process. Syst.}, vol.~31, pp. 7167--7177, 2018.

\end{thebibliography}
\end{document}